\documentclass[tcfd]{svjour}
\usepackage{times}
\usepackage{graphics}
\usepackage{amsmath}
\usepackage{amsfonts}
\usepackage{amssymb}
\usepackage{xcolor}
\usepackage[symbol]{footmisc}

\usepackage{algpseudocode}
\usepackage{algorithm}

\begin{document}
\title{Solutions to aliasing in time-resolved flow data}
%\subtitle{Do you have a subtitle?} % Insert a subtitle or remove this line
%
\Author{Ugur Karban\thanks{E-mail address of the corresponding author: ukarban@metu.edu.tr}}{D\'{e}partement Fluides, Thermique, Combustion, Institut PPrime, CNRS–Université de Poitiers–ENSMA, Poitiers, France\\
Department of Aerospace Engineering, Middle East Technical University, 06800, Ankara, Turkey}
%\Author{Eduardo Martini}{Divis\~{a}o de Engenharia Aeroespacial Instituto Tecnológico de Aeronáutica S\~{a}o Jos\'{e} dos Campos, SP, Brazil
% sample for a "present address"
%} % this closes the address field of the second author
\Author{Eduardo Martini, Peter Jordan}{D\'{e}partement Fluides, Thermique, Combustion, Institut PPrime, CNRS–Université de Poitiers–ENSMA, Poitiers, France}
\Author{Guillaume A. Br\`{e}s}{Cascade Technologies Inc., Palo Alto, CA 94303, USA}
\Author{Aaron Towne}{Department of Mechanical Engineering, University of Michigan, Ann Arbor, MI, USA}
%\Author{Andr\'{e} V. G. Cavalieri}{Divis\~{a}o de Engenharia Aeroespacial Instituto Tecnológico de Aeronáutica S\~{a}o Jos\'{e} dos Campos, SP, Brazil}
% etc
%
\commun{Communicated by }
% Will be entered by Springer
%
\date{Received date and accepted date}
\abstract{
%Avoiding aliasing in time-resolved flow data obtained through high fidelity simulations while keeping the computational and storage costs at acceptable levels is often a challenge. This paper provides a set of strategies for identifying and mitigating aliasing that are applicable even to large data sets. The approaches are illustrated on a large-eddy simulation (LES) database for a subsonic turbulent jet flow where part of the advanced analysis is aimed at calculating the forcing terms within the resolvent framework. Well-established solutions such as increasing the sampling rate or low-pass filtering to reduce aliasing are shown to be computationally very expensive for such large datasets. Alternative solutions are suggested to avoid aliasing when generating the database or to reduce it in an existing database. These approaches are tested using a non-linear Ginzburg-Landau system. The methods to eliminate aliasing make use of the access to time-derivative information and to the entire flow domain. 
Avoiding aliasing in time-resolved flow data obtained through high fidelity simulations while keeping the computational and storage costs at acceptable levels is often a challenge.  Well-established solutions such as increasing the sampling rate or low-pass filtering to reduce aliasing can be prohibitively expensive for large data sets.  This paper provides a set of alternative strategies for identifying and mitigating aliasing that are applicable even to large data sets.  We show how time-derivative data, which can be obtained directly from the governing equations, can be used to detect aliasing and to turn the ill-posed problem of removing aliasing from data into a well-posed problem, yielding a prediction of the true spectrum.  Similarly, we show how spatial filtering can be used to remove aliasing for convective systems.  We also propose strategies to prevent aliasing when generating a database, including a method tailored for computing nonlinear forcing terms that arise within the resolvent framework.  These methods are demonstrated using a non-linear Ginzburg-Landau model and large-eddy simulation (LES) data for a subsonic turbulent jet.
}
\authorrunning{Ugur Karban et al.}
\titlerunning{Solutions to aliasing in time-resolved flow data}
% See tcfdguide to enter the correct version
\maketitle
\section{Introduction} \label{sec:intro}
Data-driven approaches are gaining popularity as a means to identify and understand different mechanisms at work in turbulent flows. A variety of spatio-temporal decomposition techniques are available that can help reveal aspects of turbulent flows not easy to observe through one-point measurements. An important example is the use of modal decomposition techniques such as proper orthogonal decomposition (POD) \citep{lumley_1967,sirovich_qam_1987}, dynamic mode decomposition (DMD) \citep{rowley_jfm_2009,schmid_jfm_2010}, or spectral proper orthogonal decomposition (SPOD) \citep{lumley_jfm_1970,picard_ijhff_2000,towne_jfm_2018} to investigate coherent structures in turbulent flows. The analysis of turbulent flows using such techniques relies on space- and often time-resolved flow data. Experimental methods are generally limited in terms of the spatio-temporal resolution they can provide, which determines the minimum size of the scales that can be investigated. Computational methods, on the other hand, allow complete access to the flow domain. The use of high fidelity computational methods such as direct numerical simulation (DNS) and large-eddy simulation (LES) has become prevalent in the last decades and is now routinely used to provide spatio-temporally resolved flow data. However, running scale-resolved numerical simulations of turbulent flows for a duration sufficient to obtain converged high-order flow statistics, typically necessary for modal decomposition methods, remains challenging, and storage requirements often exceed practical limits in today's computers. A common practice to limit storage size is to downsample the simulation data. There exists, on the other hand, a maximum limit for downsampling ratio when spectral analysis is of interest: {undersampling a signal, i.e., sampling it at a rate lower than the Nyquist limit, leads to aliasing \citep{nyquist_ieee_1928,shannon_ieee_1948}.} 

An established solution to the aliasing problem involves low-pass filtering the data to satisfy the Nyquist criterion \citep{nyquist_ieee_1928,shannon_ieee_1948}. However, given the broadband nature of turbulence and the very large databases involved, low-pass filtering a numerical database can often be impracticable even with on-the-fly implementations, as is illustrated in the subsequent sections. Therefore, for many flow cases of scientific or industrial interest, generating a non-aliased, time- and scale-resolved dataset is a non-trivial task. 

{Aliasing has been extensively studied in the field of computer graphics. Most techniques in computer graphics aim to eliminate spatial aliasing in an image at a given instant in time. Some of the techniques adopted in computer graphics usually consists of spatial anti-aliasing methods, such as supersampling  and multisample anti-aliasing . In supersampling \citep{crow_ieee_1981,korein_siggraph_1983}, the image is first sampled with a resolution higher than what is necessary, and the samples falling into the same pixel in the final image are averaged. In multisample anti-aliasing \citep{akeley_siggraph_1993}, instead of sampling for every pixel in the image, neighbouring pixels are treated together to reduce the total number of samples. So called temporal anti-aliasing methods, which target spatial aliasing despite their name, use the time history of images from a video to reduce aliasing in an undersampled frame at a given instant  (see \cite{yang_cgf_2020} for a recent review). }

{Another field where aliasing is important is audio-signal processing, particularly when memoryless nonlinear transformations are applied on a digital signal to obtain some musical effects. A recent study by \cite{parker_dafx_2016} introduced the concept of antiderivative anti-aliasing, in which the integral of the nonlinear function to be applied is applied to the signal, followed by a numerical derivative to recover the signal. \cite{bilbao_ieee_2017} then extended this study to perform anti-aliasing using higher orders of integration. \cite{laPastina_dafx_2021} showed that integration can be considered as an finite impulse response (FIR) filter and extended this anti-aliasing concept to use infinite impulse response (IIR) filters instead of anti-derivatives.}

{In fluid mechanics, spatial aliasing is a well-known source of error within numerical simulations \citep{phillips_1954,orszag_jas_1971,rogallo_arfm_1984,
ghosal_jcp_1996,chow_jcp_2003}, particularly when using spectral or high-order finite-difference schemes for direct numerical simulations (DNS) or large-eddy simulations (LES). Solutions to this type of aliasing includes the famous 3/2 padding rule \citep{orszag_jas_1971,patterson_pof_1971} and random-phase-shift method \citep{rogallo_nasa_1977,rogallo_nasa_1981} for Fourier discretisations. \cite{kirby_jcp_2003} extended the 3/2 padding rule to be used in polynomial spectral methods with quadratic nonlinearities as in incompressible flows and 2/1 padding for cubic nonlinearities as in compressible flows. Alternative dealiasing methods based on splitting the convective terms during discretisation have also been proposed \citep[among many others]{kennedy_jcp_2008,winters_jcp_2018}.}

Investigation of turbulent flows via linearisation of the Navier-Stokes (N-S) equations around the mean has gained significant popularity in the last decade. A common approach is called resolvent analysis { \citep{farrel_pof_1993,jovanovic_jfm_2005,schmid_arfm_2007, mckeon_jfm_2010,hwang_jfm_2010,sipp_tcfd_2012,towne_jfm_2018,
schmidt_jfm_2018,cavalieri_amr_2019,lesshafft_prf_2019},} where the linearised N-S equations are organised in the frequency domain in input-output form. The input, i.e., the forcing, is connected to the output, i.e., the response, via a mean-flow-based linear operator, i.e., the resolvent operator. There exists a broad literature discussing the amplification mechanisms embedded in the resolvent operator \citep[among many others]{mckeon_jfm_2010,hwang_jfm_2010,beneddine_jfm_2016}
%, and leveraging them for flow estimation and control \citep(martini_jfm_2020,martini_jfm_2022,simon_prf_2019}
 and its association with the coherent structures seen in turbulent flows \citep[etc.]{schmidt_jfm_2018,towne_jfm_2018,pickering_jfm_2020}. It has been shown that, for certain cases, inclusion of the forcing term in the resolvent analysis is necessary to obtain a quantitatively accurate model of turbulent flows \citep{zare_jfm_2017,karban_jfm_2020,martini_jfm_2020,
nogueira_jfm_2021,morra_jfm_2021,karban_jfm_2022}. For this, one needs time-resolved forcing data along with the state. The storage and computational limitations arising for the state also apply for the forcing, making the aliasing a pertinent issue for the forcing as well. 

{Although there is a broad literature on aliasing related to numerical schemes, aliasing due to temporal/spatial downsampling has received considerably less attention. In this study, we aim to bridge this gap, focusing on aliasing encountered while storing DNS or LES datasets.} We investigate the applicability of some existing signal-processing techniques and offer new solutions to detect or minimise aliasing in data produced by any time-resolved simulation. We show that making use of the governing equations or, when applicable, the convection nature of the flow, it is possible to provide tailored anti- or de-aliasing solutions for the state and/or the forcing data for a given flow. Here, by anti-aliasing we mean measures to avoid aliasing while generating a database, and by de-aliasing, efforts to reduce aliasing in previously downsampled data. 

{The organization of the paper is as follows. {A brief review of the mathematical definition of aliasing together with some established solutions to detect or minimise aliasing is provided in \S2. The solutions we offer for detection, de-aliasing and anti-aliasing in dynamical systems are also given in this section.}  {A nonlinear Ginzburg-Landau model problem is used in \S4 to test the de-aliasing and anti-aliasing strategies. An LES database designed for resolvent-based prediction of flow structures in a turbulent jet is used as a case study in \S5.} Finally, some concluding remarks are presented in \S6.}

\section{Aliasing} \label{sec:aliasing}
In this section, we will provide a brief description of the phenomenon of aliasing. We will then discuss strategies to detect aliasing in a given database. Finally, we will discuss methods that can be used prior to downsampling the database to minimise aliasing or to de-alias a pre-existing database. 
\subsection{Mathematical description}
The Nyquist theorem indicates that to be able to correctly represent the frequency spectrum of a real-valued signal, the sampling frequency $f_s$ of the signal should satisfy
\begin{equation}\label{eq:NyqCrit}
f_s>2f_{max},
\end{equation}
where $f_{max}$ is the maximum frequency contained in the signal \citep{nyquist_ieee_1928,shannon_ieee_1948}. Violation of the Nyquist criterion causes an unknown bias in the power spectral density (PSD) of the signal, namely, \emph{aliasing}, which can be formalised as follows. 

Given a signal $y$ with Fourier transform (FT) $\hat{y}$, the discrete-time FT (DTFT), i.e., the FT of the sampled signal $y_n\triangleq y(n\delta t)$, where $n=1,2,\cdots$ and $\delta t=1/f_s$, is 
\begin{align}
\mathcal{F}(y_n)(\omega)=\sum_{j=-\infty}^{\infty}\hat{y}(\omega-j\omega_s),
\end{align}
where $\omega=2\pi f$ and $\omega_s=2\pi f_s$. For real-valued signals, there exists a symmetry in the frequency spectrum such that $\hat{y}(-\omega)=\hat{y}(\omega)^*$, where the superscript $*$ denotes complex conjugation. In that case, a signal satisfying the criterion given in  \eqref{eq:NyqCrit} also satisfies
\begin{align} \label{eq:aliasPart}
\sum_{\substack{j=-\infty,\\j\neq0}}^{\infty}\hat{y}(\omega-j\omega_s)=0, \textrm{ for } |\omega|<\omega_s/2,
\end{align}
which leads to $\mathcal{F}(y_n)(\omega)=\hat{y}(\omega)$. If the Nyquist criterion is not satisfied, then $\mathcal{F}(y_n)(\omega)$ provides an approximation of the spectrum, related to the true spectrum as
\begin{align} \label{eq:predSpec}
\hat{y}_p(\omega)&=\hat{y}(\omega)+\hat{a}(\omega),
\end{align}
where $\hat{a}(\omega)$ denotes the aliasing terms
\begin{align} \label{eq:aldef}
\hat{a}(\omega)&=\sum_{k=1}^{\infty}\hat{y}(\omega_{a_k}),
\end{align}
where $\omega_{a_k}=\omega_{a_{k-1}}+k(-1)^k\omega_s$ denotes the aliasing frequencies in order of appearance with $\omega_{a_0}=\omega$. In the case of monotonic decay of the spectrum beyond the Nyquist limit, the Fourier amplitudes ${\omega}_{a_k}$ decrease as $k$ increases. For smooth functions, the spectra is always monotonically decreasing for sufficiently high frequencies, and thus aliasing effects can be made arbitrarily small by increasing the sampling frequency. %In such a case, aliasing becomes less marked as one moves from the Nyquist limit to lower frequencies.  
However, for broadband signals, like turbulent quantities, the sampling frequency required to achieve acceptable levels of aliasing can be practically unattainable.

\subsection{Detecting temporal aliasing in a flow database} \label{subsec:detect}
Quantifying aliasing in a database is crucial to be able to determine what analysis, and with how much reliability, can be performed with it. However, methods to quantify aliasing are usually limited to provide an upper bound. {An easy way to check aliasing while sampling a signal is to double the sampling frequency and compare the resulting spectra with that of the original signal. {If the two spectra differ, this indicates that the original signal is undersampled, and the difference in the spectra constitutes an aliasing upper bound.} In the case of a numerical database, one may probe the data at certain positions with higher sampling rate to check the convergence of the spectra at these positions to gain an overall idea of the aliasing in the database. However, this requires prior knowledge of critical regions for aliasing and provides limited information about the spatial distribution of aliasing. In case critical regions are not known a priori or it is desired to predict aliasing in the entire domain, one needs better suited tools.} {In the following subsections, we discuss how to infer aliasing by inspecting the spectrum of a signal and introduce a new strategy based on time-derivative information to predict aliasing.}

\subsubsection{Inspecting the decay rate} \label{subsubsec:inspect}
Given a signal whose approximated spectrum is defined as $\hat{y}_p$, assumed to decay monotonically beyond the Nyquist limit $\omega_s/2$, the ratio 
\begin{align} \label{eq:aliasrat}
\left|\frac{\hat{y}_{p}(\omega_s/2)}{\hat{y}_{p,\text{min}}}\right|,
\end{align}
where $\hat{y}_{p,\text{min}}$ denotes the minimum of $\hat{y}_{p}$ within the frequency range of interest, can be used to estimate the upper bound on the first aliasing term given as $\omega_{a_1}=\omega-\omega_s$ with $\omega<\omega_s/2$. Assuming monotonic decay beyond the Nyquist limit leads to 
\begin{align}
\left|\hat{y}_p(\omega_{a_1})\right|< \left|\hat{y}_p(\omega_{s}/2)\right|,
\end{align}
which provides an upper bound on the first, and thus, dominant aliasing term. A sufficiently small ratio implies un-aliased data. 

\subsubsection{Using time-derivative data} 
{The decay-rate approach can be too conservative, and does not provide strategies to de-alias the data, if necessary. Aiming to reduce these restrictions, we propose here an alternative methodology for quantifying aliasing is outlined below. This assumes that both the state and its time derivative are available. Although, storing both the state and its time derivative is not a common practice for many applications, for dynamical systems, the time-derivative information can be obtained using the governing differential equations given access to full state.} 

Given a signal $y$ and its exact FT $\hat{y}$, the FT of $z(t)\triangleq\partial y(t)/\partial t$ is 
\begin{align} \label{eq:timeDer}
\hat{z} = i\omega\hat{y}.
\end{align}
Assuming that both $y(t)$ and $z(t)$ are available, one can obtain the FT of $z$ by calculating $\mathcal{F}(z)$ directly, or equivalently, by using  \eqref{eq:timeDer}. We will refer to the former as the `time-domain' approach and the latter as the `frequency-domain' approach. %If the approximated spectrum of $y$ contains aliasing as per  \eqref{eq:predSpec}, that of $z$ will also be aliased. 
The difference between these two approaches can be explored to estimate the aliasing present in the data. Using the time domain approach, we write
\begin{align}\label{eq:timDom}
\hat{z}_p^{(t)}(\omega) &= \hat{z}(\omega) + \sum_{k=1}^{\infty}\hat{z}\left(\omega_{a_k}\right). 
\end{align}
Given  \eqref{eq:timeDer},  \eqref{eq:timDom} can be re-written as
\begin{align} \label{eq:timAp}
\hat{z}_p^{(t)}(\omega) &= i\omega\hat{y}(\omega) + \sum_{k=1}^{\infty}i\omega_{a_k}\hat{y}\left(\omega_{a_k}\right).
\end{align}
On the other hand, the frequency-domain approach gives
\begin{align} \label{eq:specAp}
\hat{z}_p^{(f)}(\omega) &=i\omega\hat{y}_p= i\omega\hat{y}(\omega) + i\omega\sum_{k=1}^{\infty}\hat{y}\left(\omega_{a_k}\right).
\end{align}
It can be seen that the aliasing terms in \eqref{eq:timAp} and \eqref{eq:specAp} differ since $\omega\neq\omega_{a_k}$. 
The difference between the two approaches can be used to quantify aliasing. Subtracting  \eqref{eq:specAp} from  \eqref{eq:timAp} and dividing the result by $i(\omega_{a_1}-\omega)=-i\omega_s$, a prediction of the aliasing in $\hat{y}_p$ is given by 
\begin{align} \label{eq:aliasing}
a_p(\omega)=\frac{i}{\omega_s}\left(\hat{z}_{p}^{(t)} - \hat{z}_{p}^{(f)}\right)\approx \sum_{k=1}^{\infty}\hat{y}(\omega_{a_k}).
\end{align}
We will refer to this method as `derivative-based aliasing prediction'. Equation \eqref{eq:aliasing} predicts the leading aliasing term accurately while overpredicting the remaining aliasing terms due to the fact that $|\omega_{a_k}|>|\omega_s|$ for any $k>1$, and higher aliasing terms are amplified by $\omega_{a_k}/\omega_s$. Therefore, it yields an upper bound on the magnitude of the aliasing. However, this estimate is expected to be more accurate when compared to the one in \S\ref{subsubsec:inspect} since it leverages the additional time-derivative information.

When performing a input-output analysis in the frequency domain, it was shown that a correction term is needed in \eqref{eq:specAp} if a windowing function is used while taking the FT \citep{martini_arxiv_2019,nogueira_jfm_2021,morra_jfm_2021}. Assuming that the same windowing function $w$ is used for both $z$ and $y$, using the chain rule for the time derivative yields
\begin{align}
\mathcal{F}(wz)\triangleq\mathcal{F}(w\partial_ty)=\mathcal{F}(\partial_twy) - \mathcal{F}\bigl((\partial_tw)y\bigr).
\end{align}
The term, $\check{y}\triangleq\mathcal{F}\bigl((\partial_tw)y\bigr)$ should be taken into account when predicting aliasing, leading to the modified expression
\begin{align} \label{eq:aliasingCorr}
a_p(\omega)=\frac{i}{\omega_s}\left(\hat{z}_{p}^{(t)} - \hat{z}_{p}^{(f)} + \check{y}\right)\approx \sum_{k=1}^{\infty}\hat{y}(\omega_{a_k}).
\end{align}

\subsection{Anti-aliasing by low-pass filtering} \label{subsec:lpf}
As mentioned in the introduction, there exist well-established solutions to aliasing such as increasing the sampling rate and/or applying a low-pass filter before sampling the signal. Filtering in a post-processing stage is impractical, as it would require the storage of spatially- and time-resolved databases, which is what downsampling strategies aim to avoid. On-the-fly implementation of these classical anti-aliasing measures, as we will demonstrate in \S\ref{subsec:minalles}, are also often impractical. {In the following, we briefly summarise alternative low-pass filters and discuss an approach to reduce the cost of low-pass filtering.}

%\subsubsection{Low-pass filtering} \label{subsubsec:lpf}
Low-pass filtering a flow database to attenuate high-frequency content prior to downsampling can be used to minimise temporal aliasing. There are fundamentally two types of filters: finite impulse response (FIR) filters and infinite impulse response (IIR) filters. IIR filters are more efficient, i.e., large attenuation can be achieved with low-order filters, however, they usually suffer from not having a linear phase delay. The FIR filters, on the other hand, are less efficient but can always be made to yield a linear phase delay, which is advantageous as it retains the signal shape in the time domain. We first focus on this type of filter.

Filter order is a critical parameter for runtime applicability of a low-pass filter to a flow database, as it dictates the number of snapshots to be held in memory while generating the database. Filter order for a Kaiser-type FIR filter \citep{kaiser_ieee_1974} can be predicted to be
\begin{align} \label{eq:kaiserfiltord}
N=\frac{A}{\Delta\omega_p/\omega_s},
\end{align}
where $A=\left(-20\log_{10}(\sqrt{d_pd_s})-13\right)/14.6$ with $d_p$ and $d_s$ being pass-band and stop-band ripple peaks, respectively, and $\Delta\omega_p$ is the pass-band width. For a fixed frequency range of interest, i.e., constant $\Delta\omega_p$, \eqref{eq:kaiserfiltord} implies $N\propto \omega_s$. The time step used while running an LES is usually much smaller than the minimum time step required for post-processing. This causes the ratio, $\Delta\omega_p/\omega_s$ to be very small, and thus yields a large filter order, $N$. 

It is possible to significantly reduce the filter order by performing the downsampling in a multi-stage process, yielding a cascaded filter \citep{shively_ieee_1975}. At each stage, the cut-off frequency is kept constant, i.e., constant $\Delta\omega_p$, while the stop-band frequency is gradually decreased, and the signal is downsampled obeying the Nyquist limit defined by corresponding stop-band frequency. In the ideal case, one can downsample the data by halving the sampling frequency after using a filter with $\Delta\omega_p^{(i)}=\omega_s^{(i)}/4$ at each stage, until reaching the frequency range of interest, $\omega_p^{(d)}$. The overall filter order can be obtained by summing the filter orders for each filter, which yields for an $n$-stage Kaiser-type filter
\begin{align} \label{eq:filtordcas}
N = A\left(\frac{n-1}{4}+\frac{1}{\Delta\omega_p^{(d)}/\omega_s^{(n)}}\right),
\end{align}
where $\omega_s^{(n)}\triangleq\omega_s/2^{(n-1)}$. {For large $\omega_s$, \eqref{eq:filtordcas} implies a reduction in the filter order by $2^{(n-1)}$.} Examples are provided in \S\S \ref{subsec:gllpf} and \ref{subsubsec:lpassfilt}, where a cascaded filter is compared to a single-stage Kaiser-type FIR filter. 

\subsection{De-aliasing methods}
{Anti-aliasing methods are not applicable for a database that has already been downsampled. In case an existing database is aliased, one needs extra information about the time history of the system to reduce aliasing. In the following, we present a new approach to extract this extra information using the time-derivative data for systems with known governing differential equations. We also discuss using the spatio-temporal correlation for convective systems for de-aliasing.}
\subsubsection{Derivative-based de-aliasing} \label{subsec:dealias}

The estimated spectra for a given signal $y$ and its time derivative $z$ were given in \eqref{eq:predSpec} and \eqref{eq:timDom}, respectively. Assuming that the first aliasing frequency is dominant, such that
\begin{align}
\sum_{k=1}^{\infty}\hat{y}(\omega_{a_k})&\approx\hat{y}(\omega_{a_1}),\label{eq:dealassump1} \\
\sum_{k=1}^{\infty}i\omega_{a_k}\hat{y}(\omega_{a_k})&\approx i\omega_{a_1}\hat{y}(\omega_{a_1})\label{eq:dealassump2}
\end{align}
is valid, \eqref{eq:predSpec} and \eqref{eq:timDom} can be rewritten as
\begin{align} \label{eq:yp}
\hat{y}_p(\omega)&=\hat{y}(\omega)+\hat{y}(\omega_{a_1}), \\
\hat{z}_p(\omega)&=i\omega\hat{y}(\omega)+i\omega_{a_1}\hat{y}(\omega_{a_1}). \label{eq:zp}
\end{align}
If time-domain information for $y$ and $z$ is available, the above equations yield a system of linear equations that can be solved for $\hat{y}$, giving
\begin{align} \label{eq:tdpred}
\hat{y}(\omega)=\frac{i\omega_{a_1}\hat{y}_p(\omega) - \hat{z}_p(\omega)}{i\omega_{a_1}-i\omega}.
\end{align}

For dynamical systems, the governing equations can usually be written as partial differential equations of the form 
\begin{align} \label{eq:ygov}
\frac{\partial y}{\partial t}=\mathcal{G}(y),
\end{align}
where $\mathcal{G}$ is a (possibly) nonlinear operator. Equation \eqref{eq:ygov} implies that the time-derivative information for the state can be calculated by applying the nonlinear operator $\mathcal{G}$ to the state data $y$, even after the data has been downsampled.

The above analysis can be extended to include higher-order aliasing terms. The analysis to obtain second-order derivative from the downsampled data is explained below as an example. Given \eqref{eq:ygov}, the second derivative in time can be written as 
\begin{align} \label{eq:2ndder}
\frac{\partial^2 {y}}{\partial t^2}=\frac{\partial \mathcal{G}({y})}{\partial t}=\frac{\partial \mathcal{G}({y})}{\partial {y}} \frac{\partial {y}}{\partial t} = \frac{\partial \mathcal{G}(y)}{\partial y} \mathcal{G}(y).
\end{align}
The right-most term in  \eqref{eq:2ndder}, if not available analytically, can be calculated using numerical differentiation, e.g.,
\begin{align}
\frac{\partial^2 y}{\partial t^2}\approx\frac{\mathcal{G}\left(y + \epsilon\mathcal{G}(y)\right) - \mathcal{G}(y)}{\epsilon}, 
\end{align}
where $\epsilon$ is a sufficiently small number. Although mathematically straightforward, extending this analysis for higher-order derivatives may be undesirable since numerical computation of higher-order derivatives can produce increasingly large errors. 

{For real signals, the spectrum is conjugate symmetric along the frequency axis, making the assumption of the first aliasing term being dominant invalid near zero frequency. At zero frequency, the first and the second aliasing terms are complex conjugate of each other, and thus of the same magnitude, invalidating the abovementioned assumption. Assuming monotonic decay beyond the Nyquist limit, the maximum difference between the first and the second aliasing terms is obtained at the Nyquist limit. Therefore, we expect the method to yield better prediction at high frequencies. Nevertheless, it is typically expected that larger aliasing  effects are found near the Nyquist limit.} %For very low frequencies, it is possible to update method by assuming two dominant aliasing terms that are complex conjugate. Equations \eqref{eq:yp} and \eqref{eq:zp} then become,
%\begin{align} \label{eq:yp2}
%\hat{y}_p(\omega)&=\hat{y}(\omega)+\left[\hat{y}(\omega_{a_1}) + \hat{y}^*(\omega_{a_1})\right], \\
%\hat{z}_p(\omega)&=i\omega\hat{y}(\omega)+i\omega_{a_1}\left[\hat{y}(\omega_{a_1}) - \hat{y}^*(\omega_{a_1})\right], \label{eq:zp2}
%\end{align}
%where the superscript * denotes the complex conjugate. Since taking the complex conjugate is an anti-linear operator, we can obtain a solution for the above system by solving the real and the imaginary parts separately, which yields,
%\begin{align}
%\hat{y}(\omega)=\frac{\Im\left(\hat{z}_p(\omega)\right)}{\omega}+i\Im\left(\hat{y}_p(\omega)\right).
%\end{align}
\subsubsection{De-aliasing via spatial filtering in convective systems} \label{subsubsec:spatfilt}
{Spatio-temporal correlation for anti-aliasing has long been used in computer graphics \citep{shinya_cgi_1993,shinya_scj_1995,sung_ieee_2002}. Another group of methods incorporating the temporal and spatial information is called time-domain anti-aliasing  \citep{nehab_eurograph_2007,scherzer_egsr_2007,yang_cgf_2020}. In computer graphics, the aim is usually to remove aliasing in a 2-D image using temporal correlations. A similar approach is also used to reduce aliasing in medical images \citep{hu_ieee_2020}. Here, we discuss how to use this concept for de-aliasing in the time domain using spatial information in a flow database.} 

In convective systems, one can observe high spatio-temporal correlation, which links the time history of the state at a given point, $\mathbf{q}(\mathbf{x}_0,t)$, to the spatial distribution of the state at a given time instant, $\mathbf{q}(\mathbf{x},t_0)$. For a database in which the domain is discretised with a sufficiently fine grid, while suffering from undersampling in time, one can deduce the missing information for the high-frequency content by investigating wavenumber-frequency spectra.  In a purely convective 1-D system, this spatio-temporal relation can be defined as
\begin{align} \label{eq:conv1dtime}
\mathbf{q}(x,t) = \mathbf{q}(x+\Delta x,t+\Delta t),
\end{align}
where $\Delta x = c\Delta t$ with $c$ being the convection velocity. Fourier transforming \eqref{eq:conv1dtime} in time and space yields,
\begin{align} \label{eq:conv1dfreq}
\hat{\mathbf{q}}(k,\omega) = \hat{\mathbf{q}}(k,\omega)e^{i(-k\Delta x + \omega\Delta t)},
\end{align}
where $k$ and $\omega$ denote the wavenumber and frequency axes, respectively. Equation \eqref{eq:conv1dfreq} implies $\omega=ck$. {In this convective system, given a cut-off frequency $\omega_0$ which satisfies $\omega_0<\omega_s/2$, one can remove the high frequency content of the state beyond this cut-off, i.e., de-alias the data, by applying a spatial filter to remove $|k|>k_0=\omega_0/c$.}

\subsection{Aliasing in nonlinear forcing terms}

Resolvent analysis treats terms that are nonlinear with respect to fluctuations to the mean flow as a forcing on the linearized equations. Identification of the critical forcing structures requires a high-fidelity database for the forcing as well
as the state. {The forcing terms are the result of triadic interactions, which distribute energy to a wider range of frequencies than the state, pushing the effective Nyquist limit to a higher frequency and thus requiring higher sampling rates to ensure negligible aliasing in the data. In the following, we provide a brief introduction to the resolvent framework and introduce a filtering strategy dedicated to minimise aliasing in the forcing.}

\subsubsection{Navier-Stokes equation in input-output form} \label{subsubsec:ns}
We consider the Navier-Stokes (N-S) equations written in matrix form,
\begin{equation} \label{eq:NSmat}
\frac{\partial \mathbf{q}}{\partial t} = \mathcal{N}(\mathbf{q}),
\end{equation}
where $\mathbf{q}$ is the state vector and $\mathcal{N}$ is the non-linear N-S operator. Applying a Reynolds decomposition to the state vector, %The state vector is defined as $\mathbf{q}=[v \: u_x \: u_r \: u_\theta \: p]^\top$, where vector elements correspond to specific volume ($=1/\rho$), streamwise, radial and azimuthal velocities, and pressure, respectively. 
%Applying Reynolds decomposition to the state vector as follows;
\begin{equation} \label{eq:reydecomp}
\mathbf{q} = \bar{\mathbf{q}} + \mathbf{q}^{\prime},
\end{equation}
and linearising the N-S equations around the mean flow $\bar{\mathbf{q}}$,  \eqref{eq:NSmat} can be re-organized as
\begin{equation} \label{eq:NScon}
\frac{\partial \mathbf{q}^\prime}{\partial t} - \mathbf{Aq}^\prime= \mathbf{f},
\end{equation}
where prime denotes fluctuation around the mean, $\mathbf{A}=\partial \mathcal{N}/\partial \mathbf{q} |_{\bar{\mathbf{q}}}$ is a linear time-invariant (LTI) operator, and $\mathbf{f}=\mathcal{N}(\mathbf{q}) - \mathbf{A}\mathbf{q}^\prime$ is what we refer to as the forcing, which includes all the non-linear terms. The resolvent form is obtained by taking the FT of  \eqref{eq:NScon}, yielding 
\begin{equation} \label{eq:nsft}
i\omega\hat{\mathbf{q}} - \mathbf{A}\hat{\mathbf{q}}= \hat{\mathbf{f}}.
\end{equation}
Note that the FT of $\mathbf{q}$ is equal to the FT of $\mathbf{q}^\prime$ for $\omega\neq0$. Equation \eqref{eq:nsft} can be written in input-output form
\begin{equation} \label{eq:resolvent}
\hat{\mathbf{q}} = \mathbf{R}\hat{\mathbf{f}},
\end{equation}
where $\mathbf{R} = (i\omega\mathbf{I} - \mathbf{A})^{-1}$ is known as the resolvent operator. This approach has been used in the modelling of numerous flows \citep{farrel_pof_1993,jovanovic_jfm_2005, mckeon_jfm_2010,hwang_jfm_2010,towne_jfm_2018,lesshafft_prf_2019}.

\subsubsection{Anti-aliasing by integration} \label{subsubsec:intFilt}

Forcing terms are obtained by the nonlinear interaction of the fluctuating state variables. For two broadband stationary signals, $y(t)$ and $q(t)$, where both are limited in frequency by lower-upper bound, $(\omega_l,\omega_u)$, i.e., $\hat{y}(\omega)=\hat{q}(\omega)=0 \textrm{ for } \omega\notin(\omega_l,\omega_u)$, the convolution theorem implies that multiplying the two signals in the time domain yields a signal which is now limited by $(\omega_l - \omega_h,\omega_u + \omega_h)$, where $\omega_h=(\omega_u - \omega_l)/2$. {Applying the same analysis in the time domain with the assumption of real signals leads to the 3/2 zero-padding rule \citep{orszag_jas_1971}}. Considering the N-S equations, for which the nonlinearity is quadratic, the above statement indicates that forcing has more energy in high frequencies compared to the state variables. This also indicates that aliasing is higher in the forcing compared to the state, and similarly, the cost of minimising aliasing is higher in the forcing compared to the state.

The mathematical description for the aliasing of a signal $y$ and its time derivative $z$ was given by \eqref{eq:aliasPart} and \eqref{eq:timDom}. Given that $|\omega_k|>|\omega|$, the ratio of the amplitude of the aliased components to those of the true spectrum is higher in $z$ than in $y$: taking the time derivative of a signal amplifies aliasing. The reverse statement also holds, i.e., aliasing is reduced by integration. We can make use of this fact to propose an approach specifically tailored for the minimisation of aliasing in the calculation of forcing terms obtained in resolvent analysis. Taking the time integral of  \eqref{eq:NScon}, we obtain
\begin{equation} \label{eq:NSconint}
\mathbf{q} - \mathbf{Aq}_{int}= \mathbf{f}_{int},
\end{equation}
where the $(\cdot)_{int}$ denotes the integrated quantity. Note that $\mathbf{A}$ remains unchanged after integration as it is an LTI operator. Taking the FT of  \eqref{eq:NSconint} yields
\begin{equation} \label{eq:nsftint}
\hat{\mathbf{q}} - \mathbf{A}\hat{\mathbf{q}}_{int}= \hat{\mathbf{f}}_{int}.
\end{equation}
The forcing in the frequency domain can be calculated using the relation $\hat{\mathbf{f}}=i\omega\hat{\mathbf{f}}_{int}$. 
Once again, in case a windowing function is used, the windowed FT of forcing is obtained as
\begin{equation} \label{eq:fbarint}
\bar{\mathbf{f}} = i\omega\bar{\mathbf{f}}_{int} - \check{\mathbf{f}}_{int}.
\end{equation}
{This approach is related to the anti-derivative anti-aliasing method \citep{parker_dafx_2016}, where the aliasing due to applying a known nonlinear function on the signal is reduced by applying the anti-derivative of the function on the signal and then taking the time derivative via a finite difference scheme. In our case, however, the nonlinearity is not known in advance as it involves multiplication of the state with itself.}

The method requires calculating $\mathbf{q}_{int}$ through numerical integration of $\mathbf{q}$ on the fly and storing both terms at the downsampling rate. The process for application to the numerical database is given in algorithm \ref{alg:1}. {
\begin{algorithm}[t]
\caption{Computing the forcing integrated over time} \label{alg:1}
\begin{algorithmic}[1]
\State Calculate the state $\mathbf{q}$ and $\mathbf{q}_{int}$ through time-resolved simulation, and store them at every ${N}^{\text{th}}$ time step.
\State Calculate and save the mean flow $\bar{\mathbf{q}}$.  
\State Calculate and save $\mathcal{N}(\bar{\mathbf{q}})$.
\State For each snapshot, calculate $\mathbf{Aq}_{int}$. %One may use numerical differentiation as 
%\begin{align}
%\mathbf{Aq}_{int}\approx\bigl(\mathcal{N}(\bar{\mathbf{q}} + \epsilon\mathbf{q}_{int}) - \mathcal{N}(\bar{\mathbf{q}})\bigr)/\epsilon,
%\end{align}
%where $\epsilon$ is a sufficiently small number. 
\State Follow \eqref{eq:NSconint}, \eqref{eq:nsftint} and \eqref{eq:fbarint} to obtain the forcing in the frequency domain.
\end{algorithmic}
\end{algorithm}
Assuming small time step during the numerical simulation, which is the case for LES and DNS, the integral of the state can be accurately computed using a simple numerical scheme such as trapezoidal rule. Computing the forcing itself rather than its integral involves computing $\partial_t \mathbf{q}$ and $\mathbf{A}\mathbf{q}^\prime$ as given in \eqref{eq:NScon}. The latter term can be computed after filtering $\mathbf{q}$ and downsampling since $\mathbf{A}$ is a linear operator. However, computing $\partial_t \mathbf{q}$ requires applying the N-S operator $\mathcal{N}$ via \eqref{eq:NSmat} (see also \S\ref{subsec:proc}). As $\mathcal{N}$ is nonlinear, this operation has to be done prior to filtering $\mathbf{q}$. {This indicates that both $\mathbf{q}$ and $\partial_t\mathbf{q}$ should be calculated and filtered on the fly, doubling the cost of filtering. In the above method, on the other hand, the $\partial_t \mathbf{q}$ term is integrated, thus filtered, analytically yielding $\mathbf{q}$, which is already available in the database. Any other filter would have to be implemented twice for the two terms on the left-hand-side of  \eqref{eq:NSconint}. Therefore, single integration yields an efficient filtering strategy in terms of computational cost.} Since integration appears as a factor of $-i/\omega$ in the frequency domain, the attenuation level is a function of frequency and can be calculated using the ratio
\begin{align} \label{eq:intatt}
|\omega_{a_1}/\omega|=|(\omega - \omega_s)/\omega|.
\end{align}
We see from \eqref{eq:intatt} that integration is effective if $\omega\ll\omega_s$. In case attenuation from a single integration is not sufficient, one can apply integration multiple times, yielding a multi-stage filter similar to the one discussed in \S\ref{subsec:lpf}. For any filter order, the proposed approach avoids saving and filtering the forcing and the state separately.

\section{Demonstration on a model Ginzburg-Landau problem} \label{subsec:minalles}

{The applicability and the effectiveness of the above strategies for detecting and mitigating aliasing will be investigated using a model problem based on the Ginzburg-Landau equation. }

\subsection{Ginzburg-Landau problem}
The Ginzburg-Landau (G-L) model is written
\begin{equation}\label{eq:GL}
\frac{\partial q}{\partial t} + U\frac{\partial q}{\partial x}-\gamma\frac{\partial^2 q}{\partial x^2}-\mu q = -\beta q|q|+f_{\text{ext}},
\end{equation}
where $U=1$ is the convection velocity, $\gamma=0.025$ is the viscosity-like coefficient, $\mu=(1-2x/L)$ is the tuning parameter defining the local/global stability of the system on a domain $x=[0,L]$ with $L=40$, and $\beta=0.1$ denotes the factor of the non-linear term on the right hand side. {Note that \eqref{eq:GL} represents a modified version of the G-L equation having a quadratic nonlinear term. The domain is discretised using Chebyshev grids with $N=128$ points. The implicit Crank-Nicholson method is used for time integration with a time step $\Delta t=0.05$. The database is stored with a sampling rate $f_s=0.4$, corresponding to a downsampling ratio of 50. A virtual Mach number, $M=0.032$ is assigned to make the sampling rate of the model problem in terms of Strouhal number equal to that of the LES database investigated in \S\ref{subsec:minalles}. The scaling yields a sampling rate of $St=12.5$, leading to a Nyquist limit of $St=6.25$. 

The spatial support and the power spectral density (PSD) of the external stochastic forcing $f_{\text{ext}}$ is shown in figure \ref{fig:extforce}. The spatial support is limited to the region $x/L<0.1$. Beyond this region, the fluctuations grow due to a convective instability of the system \citep{huerre2000}. The PSD of the external forcing peaks around $St=10$ to enhance the high-frequency content of the response $q$, and the nonlinear term $f_q\triangleq -\beta q|q|$, ensuring non-negligible aliasing in both data when downsampled at $St=12.5$ similar to the LES case. The PSDs of the corresponding non-linear term, $f_q$ and response, $q$ at $x/L=0.5$ is shown in figure \ref{fig:fnlq} before and after downsampling. It can be seen that the true spectra for both $f_q$ and $q$ peak near the Nyquist limit (for the downsampled database), which causes significant aliasing after downsampling, reaching up to 5 dB towards low frequencies.

\begin{figure} [tb]
  \centerline{\resizebox{0.9\textwidth}{!}{\includegraphics{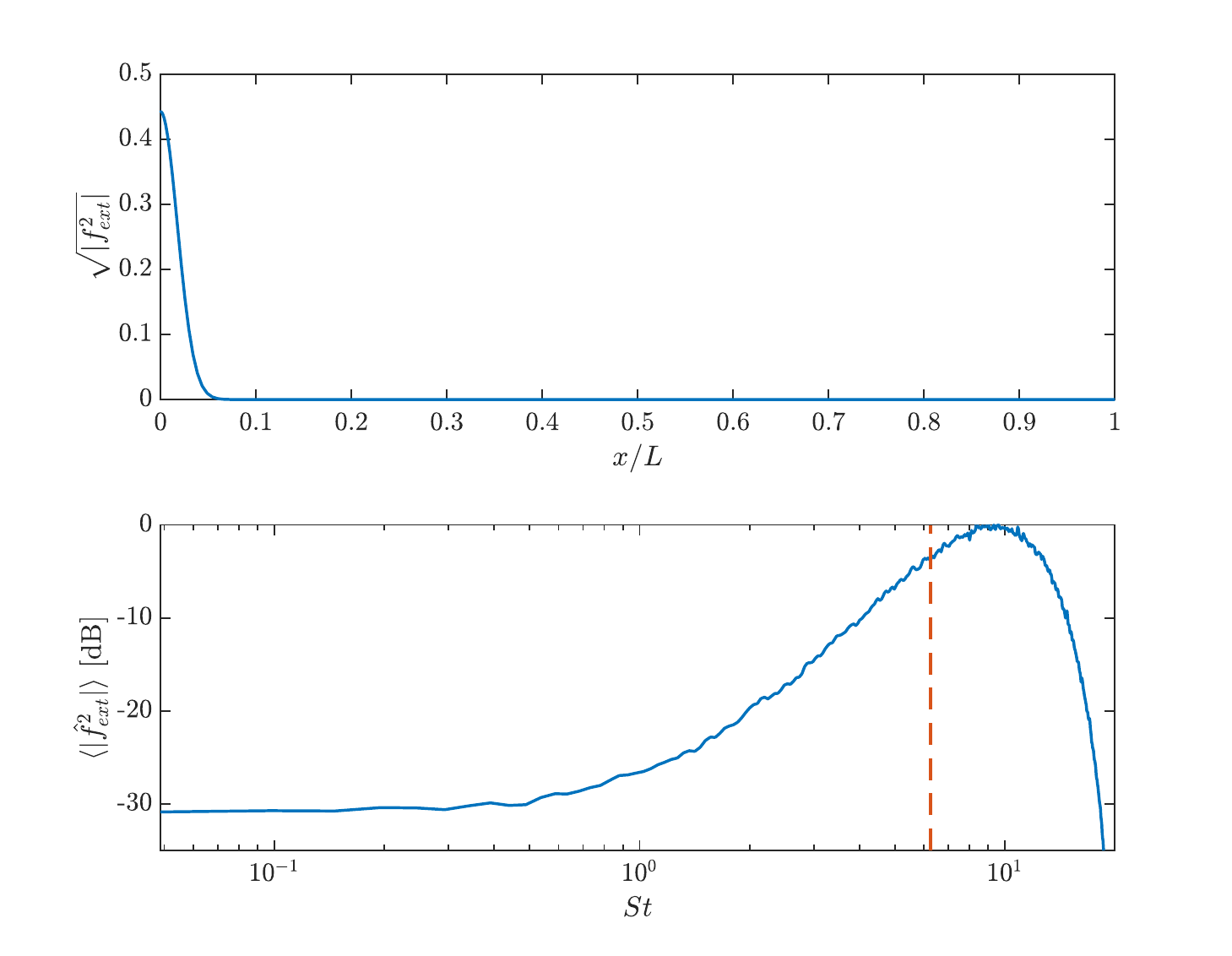}}}% Images in 100% size
  \caption{The spatial support and the spectral content of the external forcing applied on G-L system. The vertical dashed line indicates the Nyquist limit in the downsampled database.}
\label{fig:extforce}
\end{figure}

\begin{figure} [tb]
  \centerline{\resizebox{0.9\textwidth}{!}{\includegraphics{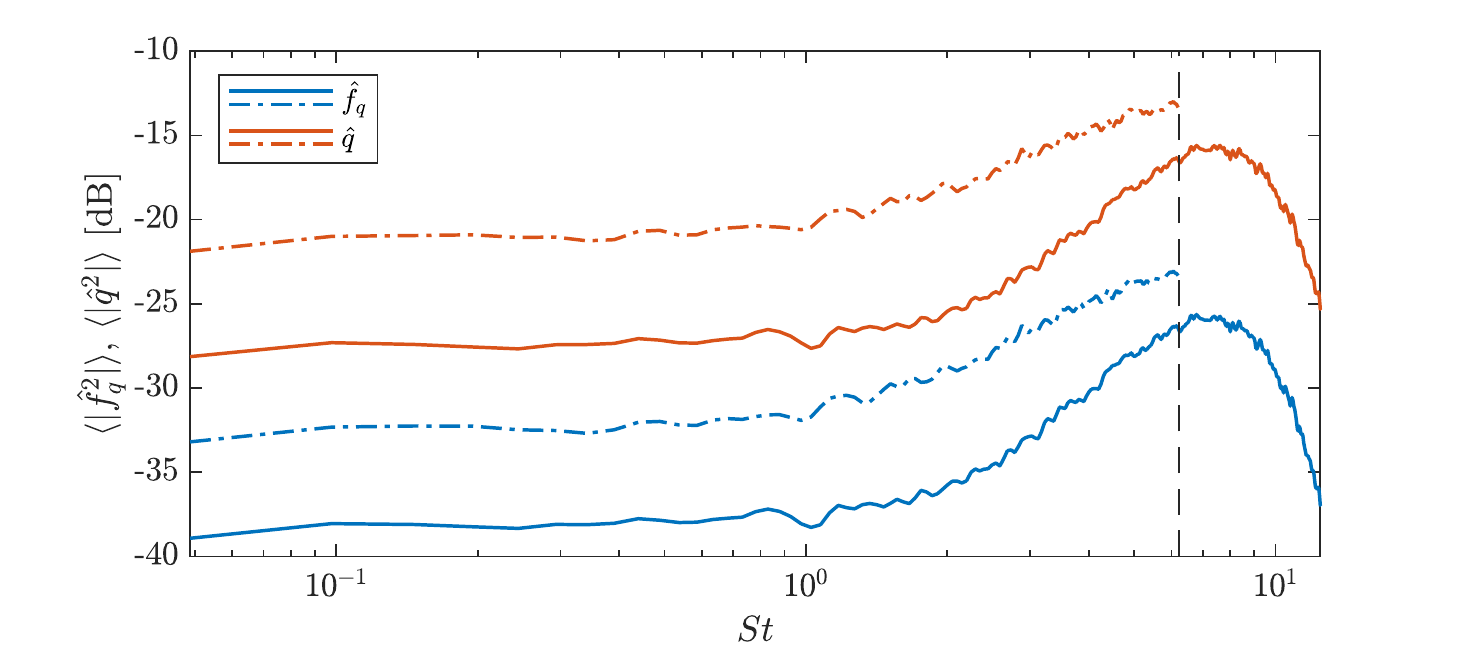}}}% Images in 100% size
  \caption{The nonlinear term, $f_q$, (blue) and the response, $q$, (orange) calculated at $x/L=0.5$. The true spectra (solid) are compared against the spectra computed after downsampling (dash-dotted). The vertical dashed line indicates the Nyquist limit of the downsampled database.}
\label{fig:fnlq}
\end{figure} 

\subsection{Detecting aliasing in the downsampled database} \label{subsec:gldetec}
{We provided a comparison of the true and the aliased spectra for the model problem in the previous subsection. In real application, one usually does not have access to high-time-resolution data to evaluate how much aliasing exists in the database. We discussed some aliasing prediction approaches in \S\ref{subsec:detect} that can be used when one has access only to the already downsampled database. We test the effectivity of these methods using the downsampled database in the model problem. The true aliasing level, $\hat{a}$, of the state is computed using the definition given in \eqref{eq:aldef} and the high-time-resolution data. The deviation from the true aliasing is measured using 
\begin{align} \label{eq:aliaslevel}
\Delta\langle|\hat{a}^2|\rangle\triangleq 10\log_{10}\langle|\hat{a}_p^2|\rangle - 10\log_{10}\langle|\hat{a}^2|\rangle,
\end{align}
where $\hat{a}_p$ denotes aliasing prediction either inspecting the decay rate or using the derivative-based method.}

{The true and the predicted aliasing maps and the differences in between are shown in figure \ref{fig:alpred}. Larger aliasing is observed towards the center of the domain and increasing frequencies. The aliasing level computed by checking the PSD at the Nyquist limit yields a constant prediction along the frequency axis and it is higher than the true aliasing upto 15 dB. Given that the difference between the true spectrum and the aliased one is around 5 dB in the entire frequency range as seen figure in \ref{fig:fnlq}, this method significantly overpredicts the aliasing in the database. Using the derivative-based method, on the other hand, one can achieve a significantly improved aliasing prediction in the entire domain. The difference between the true aliasing remains within $1$ dB. The time-derivative data is obtained using the full state data and the governing equation \eqref{eq:GL}. In case of access to full state data of dynamical systems, this method proves more useful than the first one in evaluating the quality of a given database in terms of aliasing. }

\begin{figure} [tb]
  \centerline{\resizebox{0.9\textwidth}{!}{\includegraphics{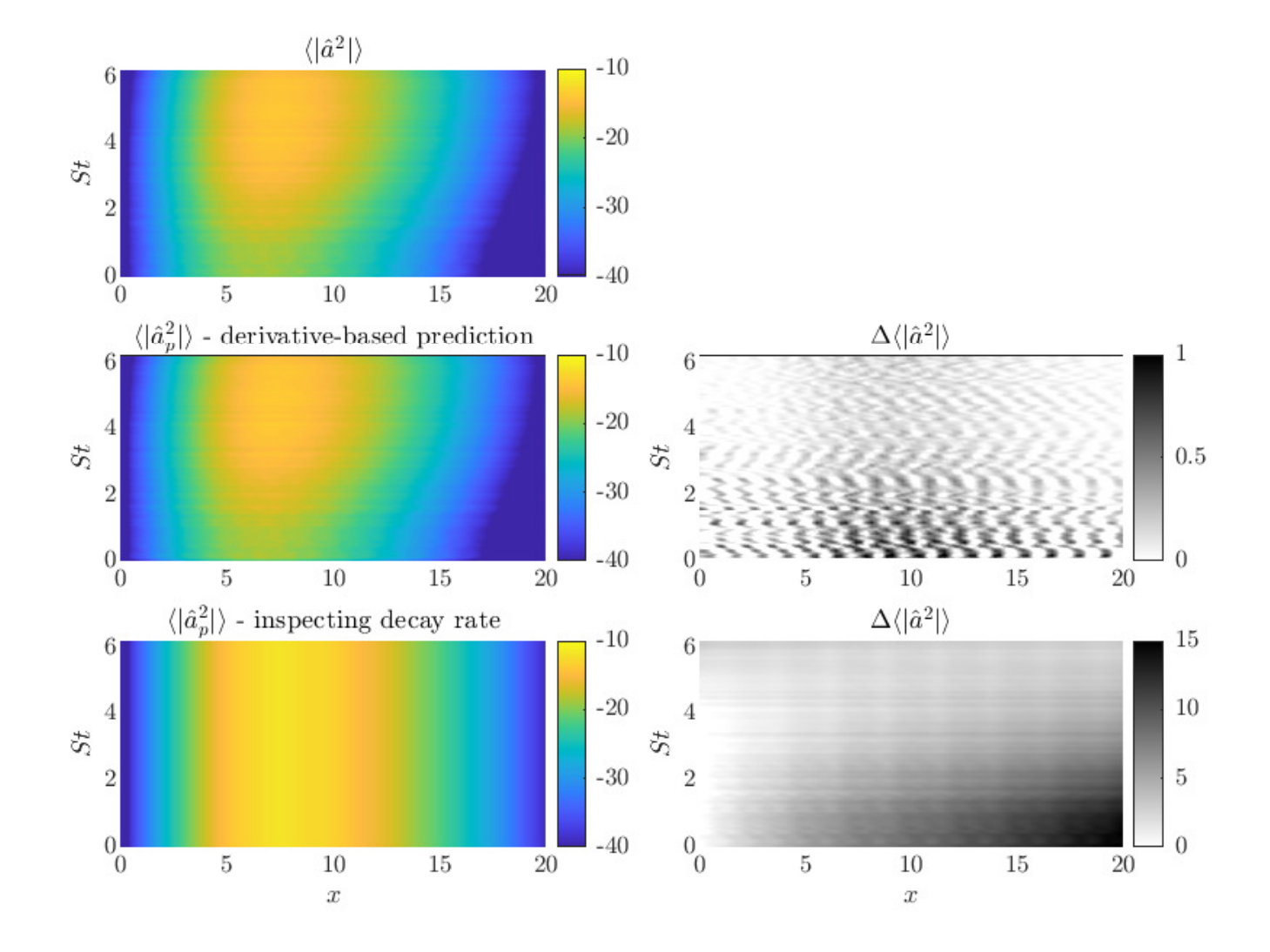}}}% Images in 100% size
  \caption{Aliasing PSD maps as a function of $x$ and $St$. Top: true aliasing; middle: derivative-based aliasing prediction (left) and its difference from the true aliasing (right); bottom: aliasing prediction by inspecting the decay rate (left) and its difference from the true aliasing (right).}
\label{fig:alpred}
\end{figure} 

\subsection{Anti-aliasing by low-pass filtering} \label{subsec:gllpf}
{The classical solution to aliasing is to low-pass filter the data prior to downsampling such that the frequency content of the signal is bounded by the Nyquist limit. The first step to design a low-pass filter (LPF) is to determine the cut-off and stop-band frequencies and the attenuation level. To avoid attenuation below the maximum frequency of interest, $St_{max}$, the cut-off frequency should be set slightly above this limit, such as 1.1$\times St_{max}$. We set the cut-off frequency as $St=4$ for the model problem. To keep the filter order at a minimum, the stop-band frequency should be equal to the Nyquist limit. The attenuation level attained at the stop-band is set as 30 dB, again, to keep the filter order at minimum while achieving adequate attenuation. We choose an FIR filter to preserve the phase relation in the data. A Kaiser type filter \citep{kaiser_ieee_1974,kaiser_ieee_1980} is designed using Matlab. The resulting filter order is found to be 462, which  determines the number of snapshots to be stored in memory while applying the filter on the high-time-resolution data.} 

{As discussed in \S\ref{subsec:lpf}, filter order can be reduced by applying a multistage filter. The Nyquist limit of the high-time-resolution data is $St=312.5$. We applied a 3-stage filter reducing the stop-band frequency from this value to the Nyquist limit of the downsampled signal, $St=6.25$. The cumulative filter order for this new filter is found to be 76, which is nearly an order of magnitude lower, and therefore, less memory intensive, than the single-stage filter. }

{We compare the aliasing level of the non-filtered data to the filtered data in figure \ref{fig:lpfpred}. The aliasing level is measured by computing the difference from the true spectrum as
\begin{align} \label{eq:aliaslevel}
\Delta\langle|\hat{q}^2|\rangle\triangleq 10\log_{10}\langle|\hat{q}_p^2|\rangle - 10\log_{10}\langle|\hat{q}^2|\rangle,
\end{align}
where $\hat{q}_p$ denotes the predicted spectrum using the downsampled database, filtered or not. Without any de-aliasing, the aliasing level is around 5 dB in the entire frequency range. Both single- and multi-stage filters yield a similar anti-aliasing performance. The aliasing is removed up to the cut-off frequency, beyond which the signal is attenuated by 30 dB before reaching the Nyquist limit.}

\begin{figure} [tb]
  \centerline{\resizebox{0.9\textwidth}{!}{\includegraphics{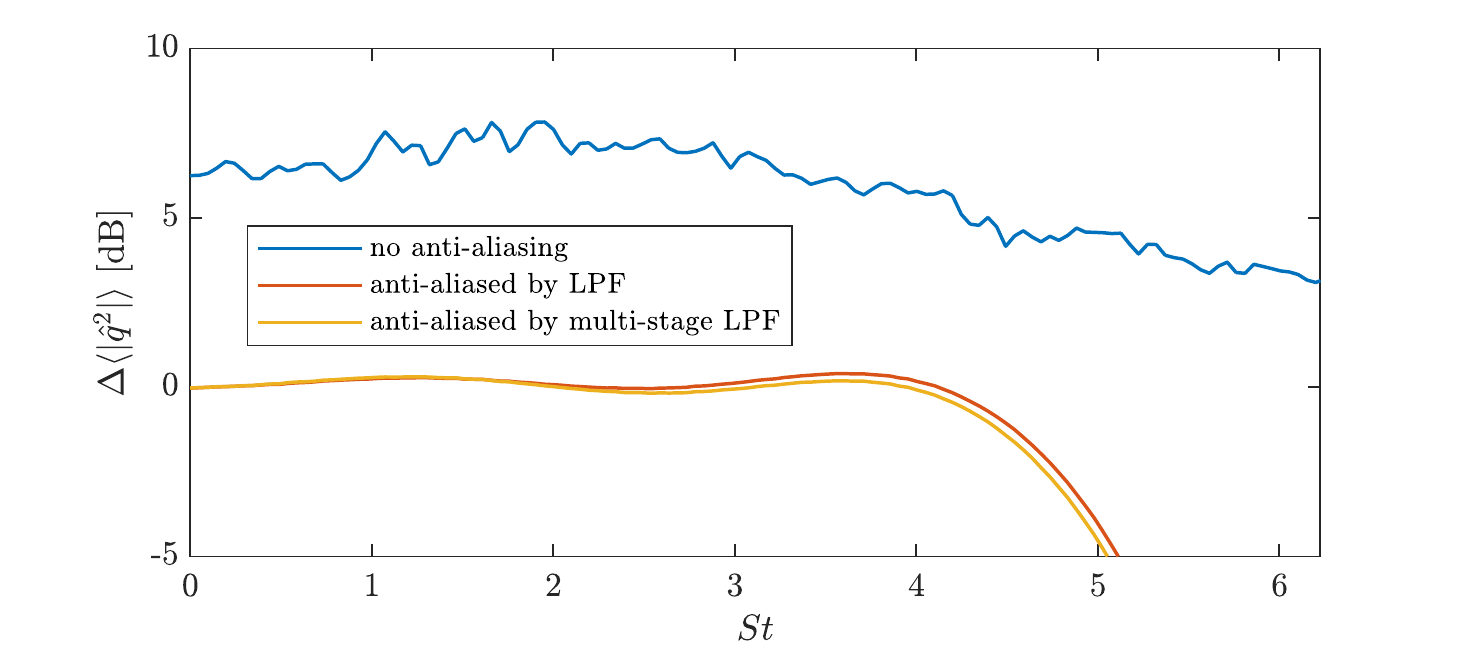}}}% Images in 100% size
  \caption{Aliasing levels measured by \eqref{eq:aliaslevel} at $x/L=0.5$ for the donwsampled $\hat{q}$ with no-filtering (blue), single-stage filtering (orange) and 3-stage filtering (yellow).}
\label{fig:lpfpred}
\end{figure} 

\subsection{Anti-aliasing by integration} 

{We discussed that integrating the linearised N-S equations as in \eqref{eq:NSconint} can be an effective way to minimise aliasing in the forcing at low frequencies. In this approach, we need to compute the integrated state during the computation of the state and both signals should be stored when downsampling. For large databases, the forcing integrated over time should be computed using the algorithm \ref{alg:1}. While alternative approaches are feasible for this simple model problem, we use this same algorithm for consistency. The deviation from the true spectrum due to aliasing when downsampled and the effect of anti-aliasing to minimize this deviation is shown in figure \ref{fig:filtcomp}. The anti-aliasing method based on integration reduces the aliasing effect in $f_q$ to less than 1 dB for the frequency range $St<2$. At higher frequencies, aliasing level slightly increase, reaching 7 dB at the Nyquist limit. The attenuation of the aliasing terms scales with $\omega_s/\omega$ as given in \eqref{eq:intatt}, leading to better performance at low frequencies but no attenuation at the Nyquist limit. The attenuation level can be enhanced using multi-stage integration. Each integration produces a geometric increase in attenuation level, proportional to $\omega_s^n/\omega^n$, where $n$ is the number of stages, with a linear increase in the memory cost proportional to $n$. Aliasing level attained when using double integration is shown in the same figure. The frequency limit for an aliasing level of 1 dB now extends to $St=4$. The anti-aliasing method fails at the frequencies very close to zero. This is mainly due to the fact that the calculation process involves division by $\omega$ due to integration, and when $\omega$ is close to zero, any error in the predicted spectrum due to DTFT is significantly amplified. However, this is an issue only for very low frequencies, and the prediction can be improved by increasing the number of snapshots when computing the DTFT.}

\begin{figure} [!htb]
  \centerline{\resizebox{0.9\textwidth}{!}{\includegraphics{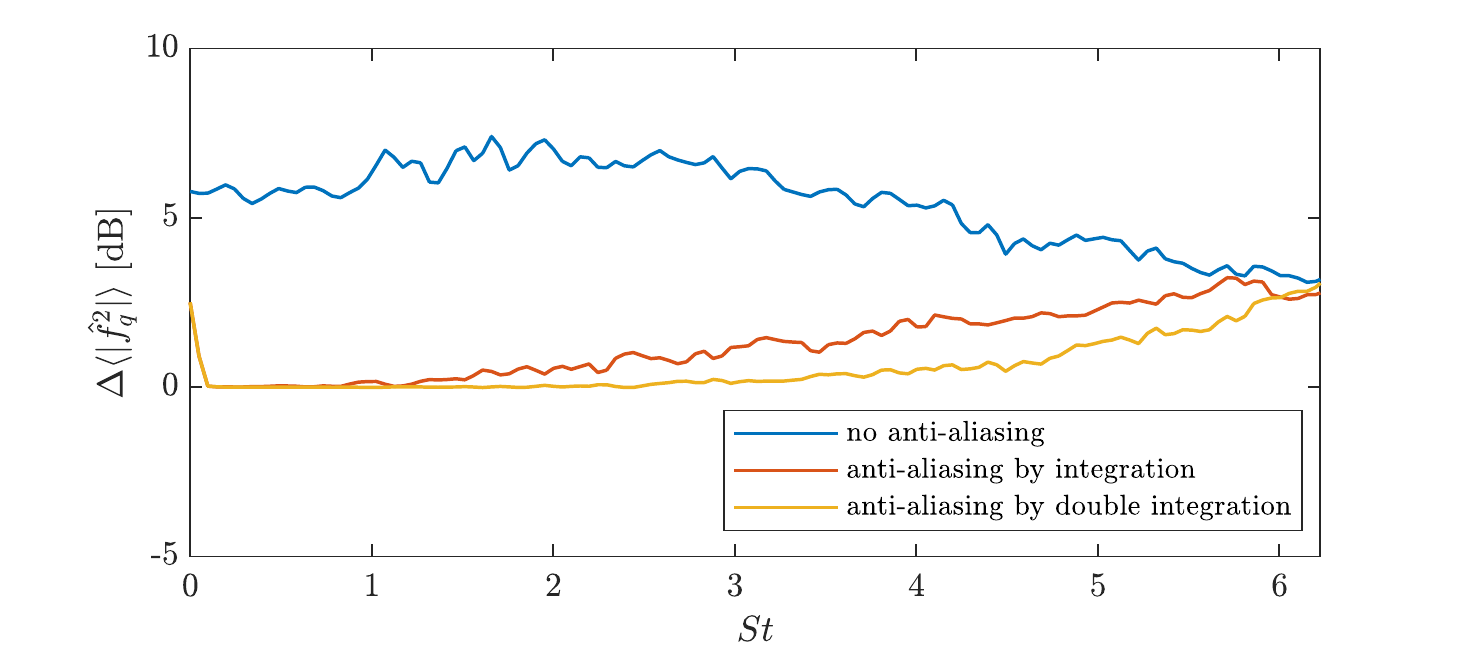}}}% Images in 100% size
  \caption{Aliasing levels at $x/L=0.5$ for the donwsampled $\hat{f_q}$ with no anti-aliasing (blue) and with anti-aliasing by integration (orange).}
\label{fig:filtcomp}
\end{figure}

\subsection{Derivative-based de-aliasing} \label{subsec:gltd}
{We showed in \S\ref{subsec:dealias} how the true spectrum of a signal can be estimated by using time-derivative information and assuming that aliasing is dominated by the leading aliasing term. This will be tested using the downsampled G-L problem. Given the time history of the state $q$ and the external forcing $f_{ext}$, the time derivative is computed using \eqref{eq:GL}, and the de-aliased spectrum is then predicted via \eqref{eq:tdpred}.  Reduction in the aliasing level for the downsampled database when applying derivative-based de-aliasing is shown in figure \ref{fig:tdpred}. The method is seen to substantially remove aliasing towards the Nyquist limit, while the error in the spectrum at frequencies close to zero exceeds the level observed in the signal with no de-aliasing. }

{The reason for the poor performance at low frequencies is the violation of the assumption that the leading aliasing term is dominant over others. Using the high-time-resolution data with sampling rate $St=312.5$, we predict the first three aliasing terms when the database is downsampled with a sampling rate $St=12.5$. The true spectrum and the aliasing terms for $q$ at $x/L=0.5$ are shown in figure \ref{fig:aliasterms}. The frequency at which the difference between the PSDs of the first and second aliasing terms reach 20 dB, indicating a threshold for the assumption of dominant first aliasing term to be valid, is marked with a vertical dashed line around $St=4$. For the frequency range to the right of this line, the de-aliasing method can be used to approximate the true spectrum. The error in the de-aliased signal is seen to rise in figure \ref{fig:tdpred} for the frequencies lower than this threshold as the dominant leading aliasing term assumption is no longer valid. Note that we present an intentionally difficult problem where the spectrum peaks at the Nyquist limit. In case the spectrum starts decaying before the Nyquist limit, even minimally, the aliasing at lower frequencies is expected to be less than that in higher frequencies.}

\begin{figure} [tb]
  \centerline{\resizebox{0.9\textwidth}{!}{\includegraphics{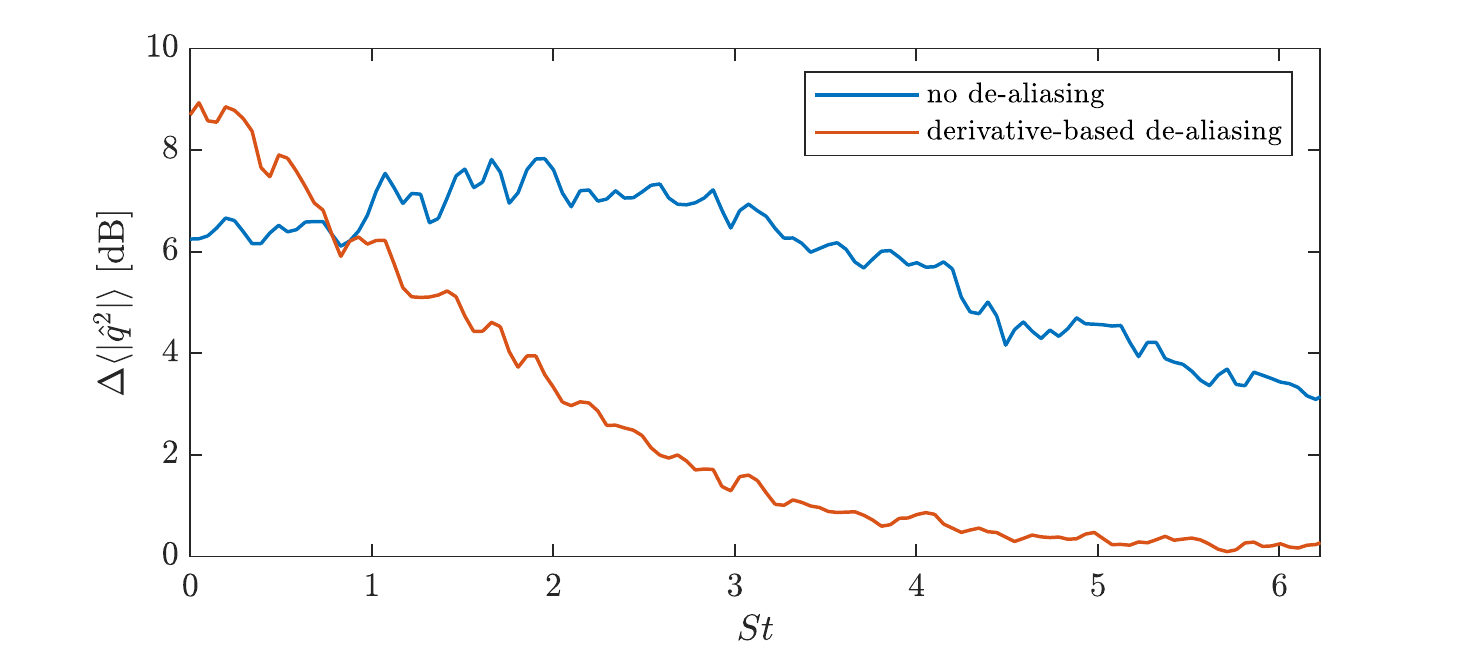}}}% Images in 100% size
  \caption{Aliasing levels measured by \eqref{eq:aliaslevel} at $x/L=0.5$ for the donwsampled $\hat{q}$ with no de-aliasing (blue) and with derivative-based de-aliasing (orange).}
\label{fig:tdpred}
\end{figure} 

\begin{figure} [tb]
  \centerline{\resizebox{0.9\textwidth}{!}{\includegraphics{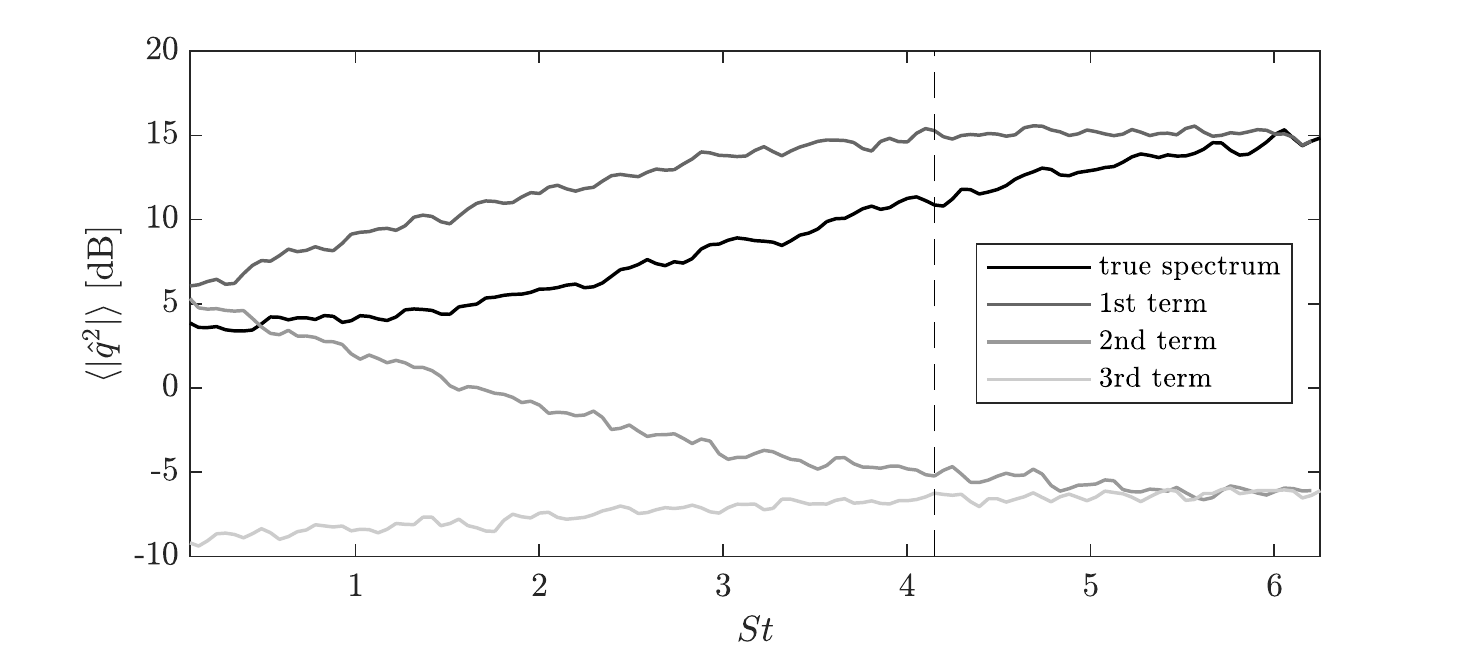}}}% Images in 100% size
  \caption{Comparison of the true PSD of $q$ at $x/L=0.5$ and the first three aliasing terms when sampled at $St=12.5$. The vertical dashed line indicates the frequency where the magnitude of the first and second aliasing terms differ by 20 dB.}
\label{fig:aliasterms}
\end{figure} 

\subsection{De-aliasing by spatial fitlering} \label{subsec:glspat}
{The Ginzburg-Landau model is a convective system. Although the nonlinear term in \eqref{eq:GL} introduces some de-correlation, the dynamics are still dominated by the convection, yielding a strong correlation for the fluctuations convected across the domain. In the model problem, we downsample the data only in time; therefore, the spatial resolution is preserved in the downsampled database. This implies that the snapshots in the downsampled database do not contain any spatial aliasing, which allows us to de-alias the database by applying a spatial filter as discussed in \S\ref{subsubsec:spatfilt}. We use this approach to de-alias the nonlinear terms in equation \eqref{eq:GL}, which are treated as an exogenous forcing term in the resolvent framework (see the discussion in \S\ref{subsubsec:ns}). Note that de-aliasing the state, if spatially well-resolved, is also possible with this approach, but not shown here to avoid repetitive analysis. }

{In order to decide the spatial filter to be applied, we convert the forcing data into the wavenumber-frequency domain by taking FT in time and space and estimating the PSD using Welch's method. The resulting spectrum is shown in the top plot of figure \ref{fig:spatfilt}. Multiple bands are observed in the spectrum, where the one in the center represents the true spectrum and the remaining ones indicate the aliasing terms. The temporal PSD can be considered as the integration of this spectrum along the wavenumber axis. Such an integration adds the true spectrum and the aliasing terms yielding the aliased spectra seen in figure \ref{fig:fnlq}. Applying a spatial filter on the data can remove the aliasing terms while mostly preserving the true spectrum. The cut-on and stop-band wavenumbers of the spatial filter to be applied are marked on the spectrum. One can see that placing the stop-band line at the beginning of the first aliasing term would cause a significant portion of the true spectrum at high frequencies to be filtered as well. Therefore a compromise is to be made between removing the aliasing terms and minimising the attenuation of the true spectrum. We used an FIR type spatial filter with 50 dB attenuation between the cut-off and stob-band wavenumbers, $k=0.2$ and 0.4, respectively. The wavenumber-frequency spectrum of the resulting filtered data is shown in the bottom plot. The aliasing for the frequencies $St<2$ is completely removed while some of portion of the leading aliasing term remained at higher frequencies. We also see that the true spectrum is partially filtered in the frequency region $St>4$.} 

{The aliasing level in the resulting filtered forcing data is calculated using \eqref{eq:aliaslevel} and compared against that of the unfiltered forcing data in figure \ref{fig:fpredspat}. The aliasing is substantially removed for $St<4$, although some attenuation reaches upto 5 dB beyond this frequency. We see that spatial filtering is an effective de-aliasing method at low frequencies in convective systems. }

\begin{figure} [tb]
  \centerline{\resizebox{0.9\textwidth}{!}{\includegraphics{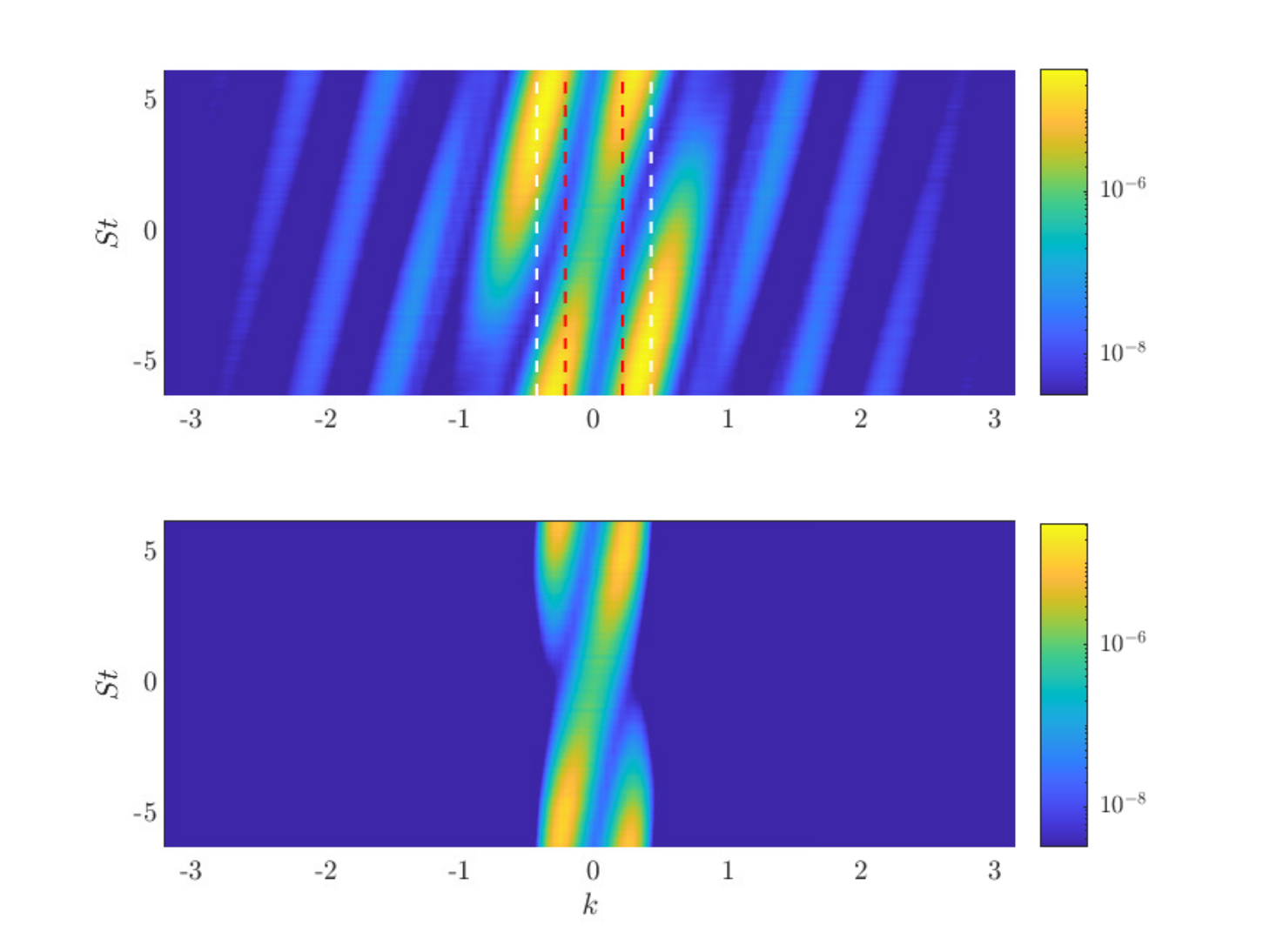}}}% Images in 100% size
  \caption{The wavenumber-frequency spectrum of the downsampled forcing $f_q$ with no filtering (top) and after applying spatial filter (bottom). Red and white dashed lines indicate the cut-off and stop-band frequencies, respectively.}
\label{fig:spatfilt}
\end{figure} 

\begin{figure} [tb]
  \centerline{\resizebox{0.9\textwidth}{!}{\includegraphics{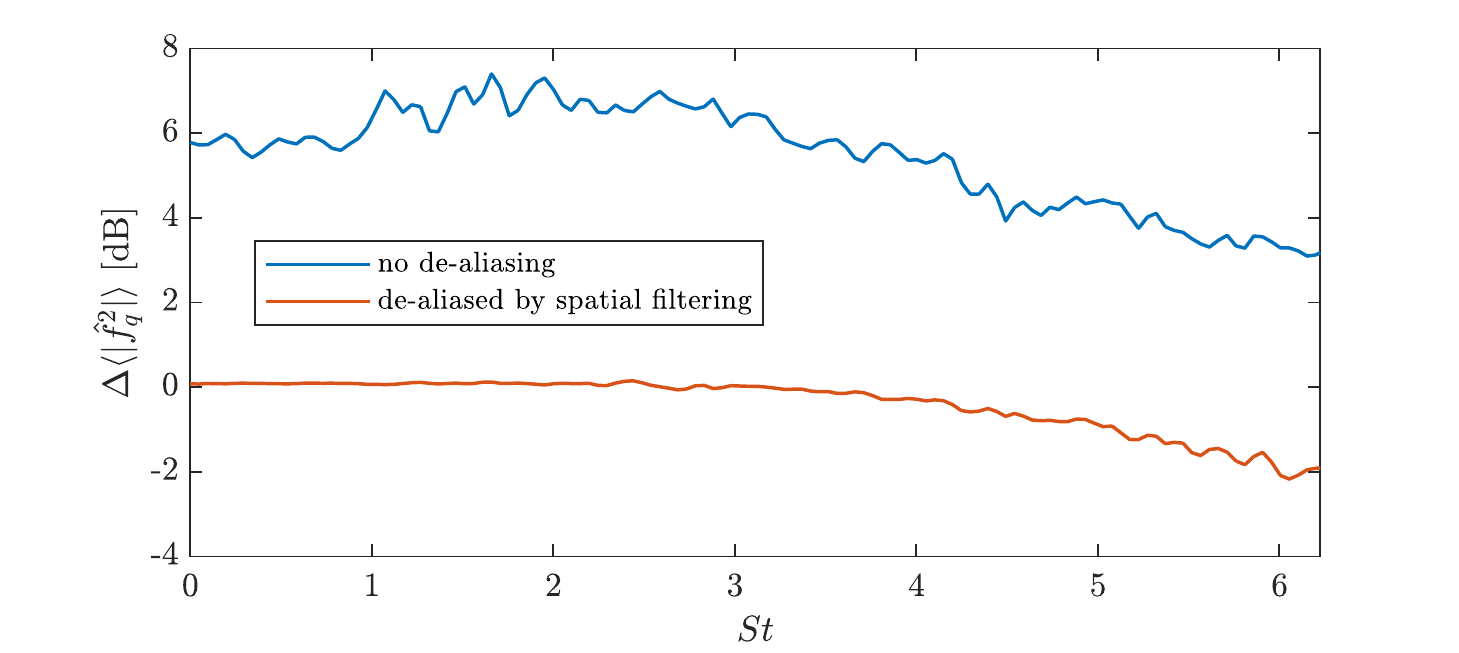}}}% Images in 100% size
  \caption{Aliasing levels at $x/L=0.5$ for the donwsampled $\hat{f_q}$ with no filtering (blue) and with spatial filtering (orange).}
\label{fig:fpredspat}
\end{figure}

\section{Application to large-eddy simulation of a turbulent jet} \label{subsec:minalles}

As discussed earlier, classical anti-aliasing solutions involve increasing the sampling rate and low-pass filtering the data prior to downsampling. The cost of these approaches when applied to an LES database of a subsonic compressible turbulent jet will be shown to be prohibitively high. The database consists of both the state and forcing data. The aliasing level in the state will be predicted using the time-derivative of state, while that in the forcing will be predicted inspecting the PSD at the Nyquist limit. Time-derivative-based methods discussed in this study are not practical to apply on the forcing since computing the time derivative of the forcing in an already downsampled database requires solving the transport equations for the nonlinear terms in the N-S equations, which can be complicated. In case of de-aliasing the state, one has the advantage of using the nonlinear operator provided in the numerical solver to obtain the time-derivative information as shown in \S\ref{subsec:proc}. We will employ the de-aliasing methods based on the time-derivative information and spatial filtering to reduce aliasing in the state and forcing, respectively.}

\subsection{Simulation and data} \label{sec:numcase}

The numerical data we use to benchmark the de-aliasing strategies was provided by LES of a $M=0.4$ jet intended for aeroacoustics analysis. The dataset has been validated against numerous experimental data for its mean statistics, boundary layer spectrum, far-field acoustic spectrum, etc. \citep{bres_aiaa_2015,bres_aiaa_2017,bres_jfm_2018}. The simulation was conducted using a grid containing 16 million elements.  Data was collected for 2000 acoustic time units ($t c_\infty / D$) with a sampling rate of $\Delta t c_\infty/ D = 0.001$, where $\tilde{t}$, $c_{\infty}$, and $D$ denote the dimensional time, speed of sound and the nozzle diameter, respectively. {A snapshot including the nozzle and the external flow domain is shown in figure \ref{fig:snapshot}. The fluctuations in temperature indicate turbulent activity, which remains within the shear layer starting from the nozzle edge and spreading in the radial direction as eddies move downstream.} %The data was then interpolated on a cylindrical grid of $(656\times138\times128)$ elements in axial, radial and azimuthal directions, respectively, at every 200 time steps ($\Delta t=0.2$). Further details about the simulation parameters can be found at \citet{bres_aiaa_2015}. 

\begin{figure} [htb]
  \centerline{\resizebox{1\textwidth}{!}{\includegraphics{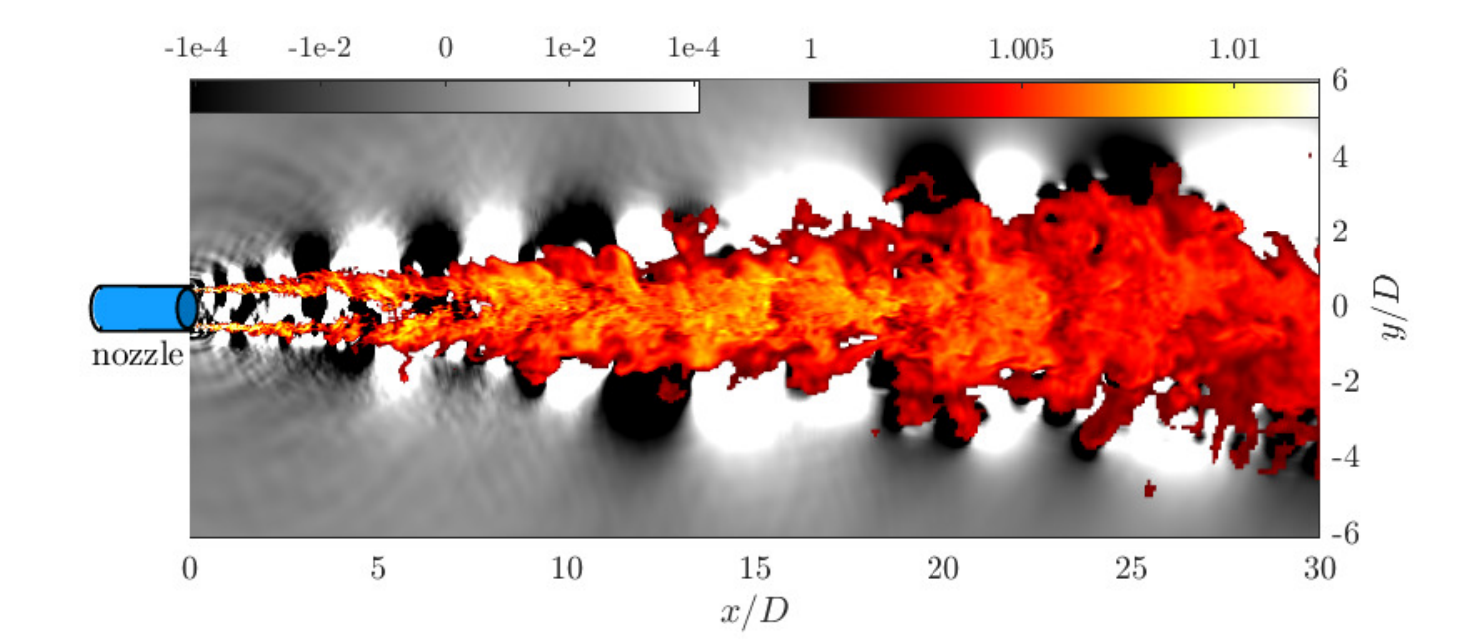}}}% Images in 100% size
  \caption{A snapshot of temperature (color) and pressure (grayscale) out of the database on structured cylindrical grid extracted from the jet LES. The nozzle is shown schematically on the left.}
\label{fig:snapshot}
\end{figure} 

The database was specifically designed for resolvent-based analysis of the jet. In the simulation, the variables $\rho$, $u_x$, $u_y$, $u_z$, and $p$, which denote the density, velocities in the $x$, $y$ and $z$ directions and pressure, respectively, are stored at runtime every 200 time steps ($\Delta t$ = 0.2) for the full computational domain. To facilitate commonly used analysis methods, the LES data is interpolated from the original unstructured LES grid onto structured cylindrical grids in the jet plume and inside the nozzle as a postprocessing step. For the jet plume,  the three-dimensional cylindrical grid extents to $0<x/D<30$, $0<r/D<6$, with $(N_x,N_r,N_\theta) = (626,138,128)$, where $N_x$. $N_r$ and $N_\theta$ are the number of points in the streamwise, radial
and azimuthal direction, respectively \citep{bres_aiaa_2016}.  Additional variables can also be computed and extracted from the LES database in order to accommodate analysis of the forcing terms that appear in resolvent analysis. {The linearisation around the mean flow, which was necessary for the resolvent analysis, was performed using a primitive-like set of variables given as, $\mathbf{q}=[\nu\,\mathbf{u}\,p]^\top$, where $\nu=1/\rho$ is the specific volume and $\mathbf{u}$ is the velocity defined in cylindrical coordinates (c.f. \cite{karban_jfm_2020}). We here briefly describe the data-extraction and -storage strategy we adopted that is associated to resolvent analysis.}

\subsubsection{Procedure to generate the LES database} \label{subsec:proc}
A complete evaluation of the linear dynamics of coherent structures requires computation of $\mathbf{f}$ and its projection into the input space of the resolvent operator $\mathbf{R}$. The LES database, thus, contains the state $\mathbf{q}$ and the two terms on the left-hand-side of  \eqref{eq:NScon}, which are used to compute the forcing term $\mathbf{f}$. These terms are computed according to the procedure given in algorithm \ref{alg:2} using the state data \citep{towne_phd_2016}. {The term $\epsilon$ used for numerical computation of the Jacobian is set as $10^{-7}$. \citet{towne_phd_2016} reported that the resulting forcing data are not affected by the value of $\epsilon$ for a wide range spanning a couple of orders of magnitude.}
{
\begin{algorithm}[t]
\caption{Computing the forcing} \label{alg:2}
\begin{algorithmic}[1]
\State Calculate the state $\mathbf{q}$ through LES with $\Delta t=0.001$ and store it at every 200${}^{\text{th}}$ time step.
\State Calculate and save the mean flow $\bar{\mathbf{q}}$.  
\State Calculate and save $\mathcal{N}(\bar{\mathbf{q}})$.
\State For each snapshot, calculate $\partial \mathbf{q}/\partial t = \mathcal{N}(\mathbf{q})$.
\State For each snapshot, calculate $\mathbf{A}\mathbf{q}^\prime\approx{\bigl(\mathcal{N}(\bar{\mathbf{q}} + \epsilon\mathbf{q}^\prime) - \mathcal{N}(\bar{\mathbf{q}})\bigr)}/{\epsilon}$, where $\epsilon$ is a sufficiently small number. 
\State Interpolate $\mathbf{q}$, $\partial \mathbf{q}/\partial t$, and $\mathbf{A}\mathbf{q}^{\prime}$ data onto the  cylindrical grid.
\State Compute the forcing in the time domain as $\mathbf{f}=\partial \mathbf{q}/\partial t-\mathbf{Aq}^{\prime}$.
\end{algorithmic}
\end{algorithm}
}

The interpolation onto a cylindrical grid is performed to facilitate an azimuthal Fourier-series expansion of the jet dynamics. Stored at single precision on the cylindrical grid, the entire LES database amounts to 6.6 TB of data.
\subsubsection{Aliasing in the LES database}

\begin{figure} [!htb]
  \centerline{\resizebox{1\textwidth}{!}{\includegraphics{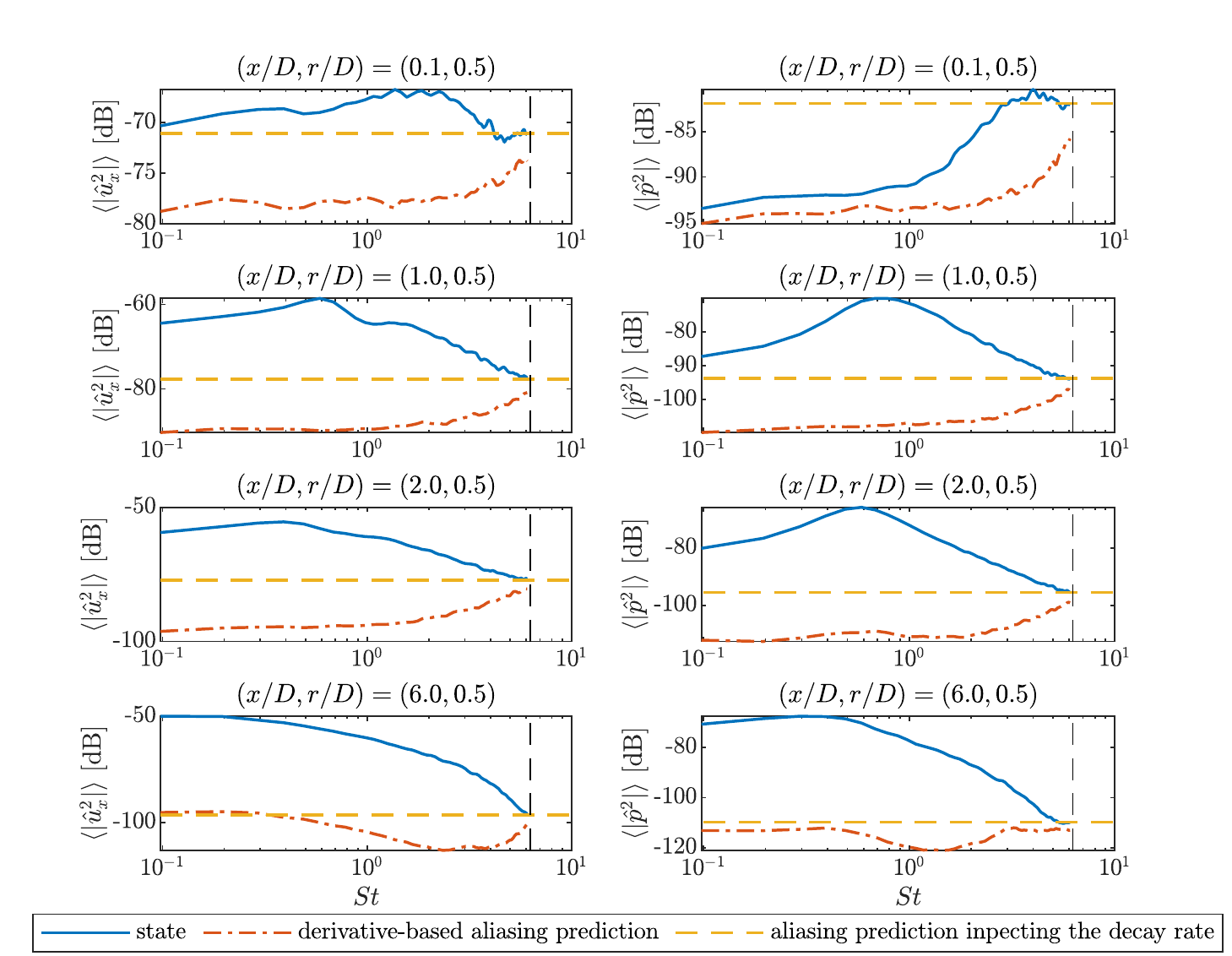}}}% Images in 100% size
  \caption{Comparison of the PSDs of $u_x$ (left) and $p$ (right) to that of the corresponding aliasing predicted by the derivative-based method. Solid: state, horizontal-dashed: aliasing prediction assuming monotonic decay, dash-dotted: derivative-based aliasing prediction, vertical-dashed: Nyquist limit. }
\label{fig:psd_dux}
\end{figure} 

\begin{figure} [htb]
  \centerline{\resizebox{0.8\textwidth}{!}{\includegraphics{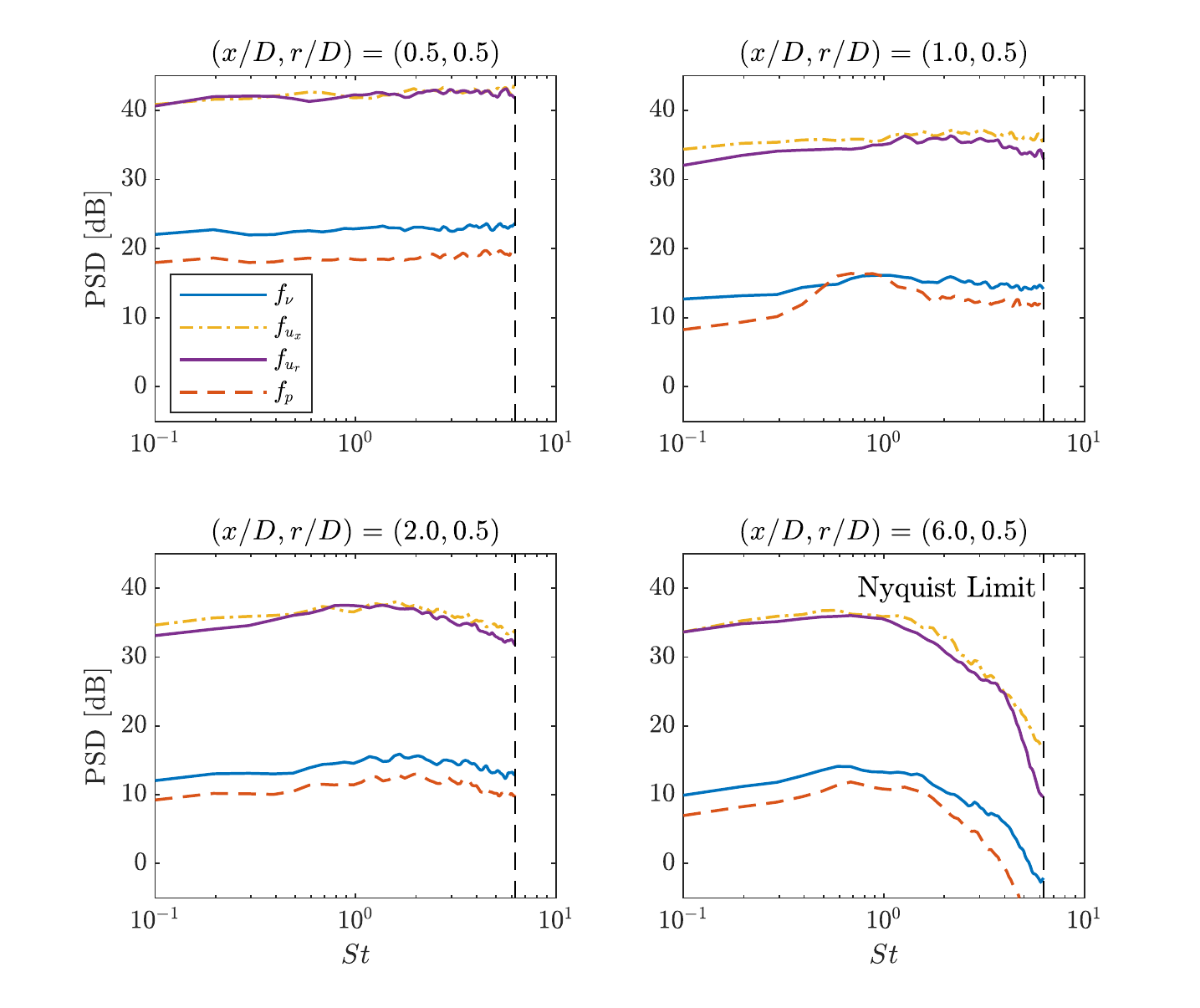}}}% Images in 100% size
  \vspace{-10px}
  \caption{PSD of the forcing terms, $f_\nu$ (solid-blue), $f_{u_x}$ (dash-dotted), $f_{u_r}$ (solid-violet), and $f_p$ (dashed) at various axial positions on the lip line.}
\label{fig:forcepsd}
\end{figure}

\begin{figure} [!htb]
  \centerline{\resizebox{.84\textwidth}{!}{\includegraphics{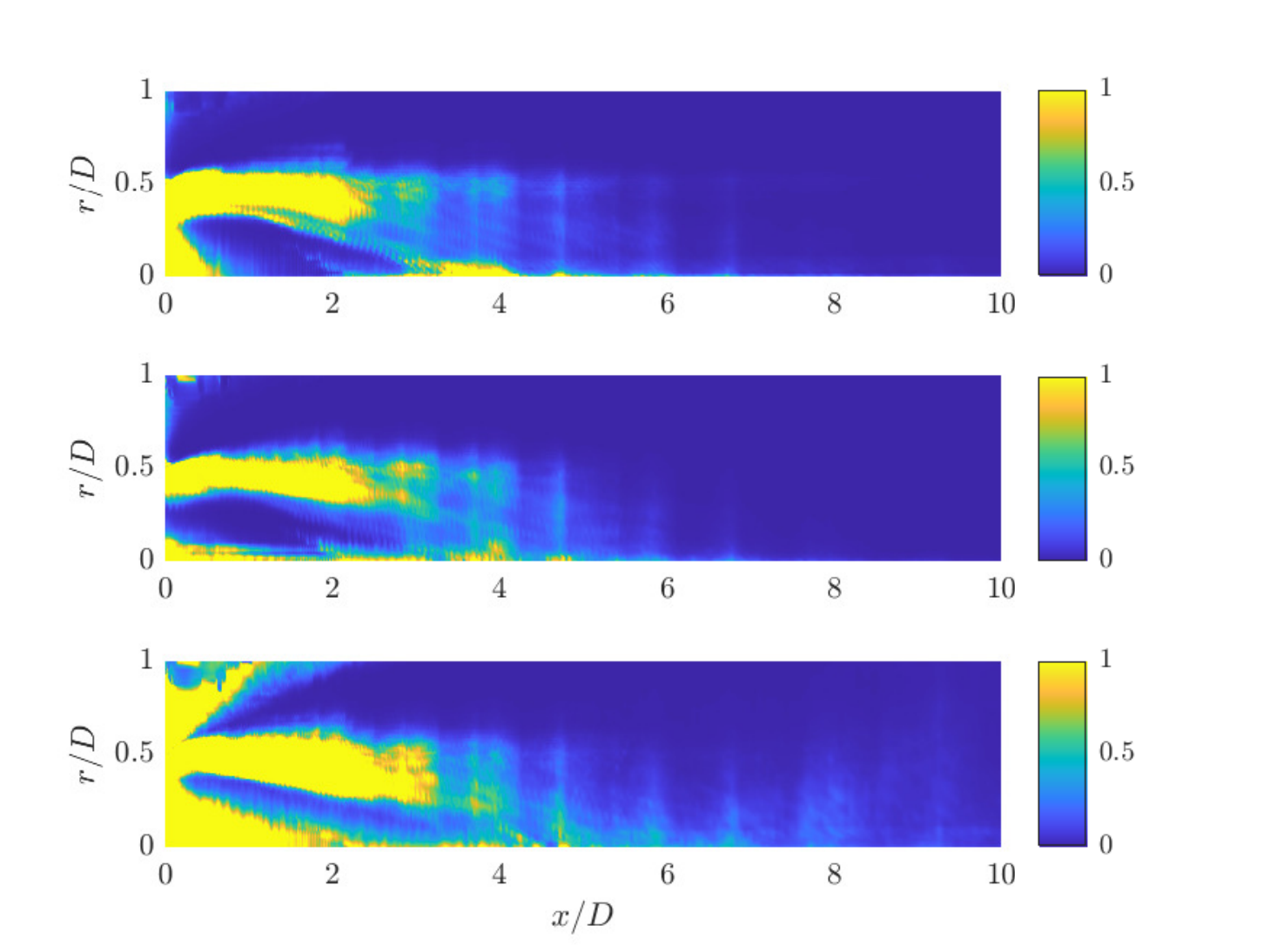}}}% Images in 100% size
  \vspace{-5px}
  \caption{Ratio of the forcing PSD at the Nyquist limit to the minimum PSD value below $St=1$ for the forcing terms, $f_{u_x}$ (top), $f_{u_r}$ (middle), and $f_p$ (bottom) to indicate the aliasing in the spectrum.}
\label{fig:forceflat}
\end{figure}

{Aliasing in the LES database can occur due to undersampling in time and/or in space. Here, we will focus on temporal aliasing, while a short discussion on spatial aliasing in the LES database is provided in the appendix.} We compare the two strategies discussed in \S\ref{subsec:detect} to predict aliasing. While taking the DTFT, an exponential windowing function
\begin{align} \label{eq:win}
W(t) = e^{n\left(4-\frac{T}{t(T-t)}\right)},
\end{align}
given in \cite{martini_arxiv_2019} is used with $n=1$ and window size $T=128\Delta t$. The PSDs are computed using Welch's method \citep{welch_ieee_1967} with a 75\% overlap between consecutive time blocks. The scope of the test case is limited to the first azimuthal Fourier mode, which can be obtained via axisymmetric averaging, for a frequency range $[0.1,1]$ given in terms of Strouhal number, $St=\tilde{f}D/U_j$, where $\tilde{f}$, $U_j$ and $D$ denote dimensional frequency, nozzle exit velocity, and nozzle diameter, respectively. The sampling frequency and Nyquist limit for this configuration are $St_s=12.5$ and $St_N=6.25$, respectively.

The PSDs of the predicted aliasing is shown in figure \ref{fig:psd_dux} in comparison to the PSD of the state for the streamwise velocity and pressure terms. The temporal resolution is insufficient in regions for which the PSD of the aliasing is comparable to or larger than the PSD of the state. Aliasing is more dominant near the nozzle, and higher for pressure than for streamwise velocity. Recall that all the prediction methods discussed above provide an upper bound for aliasing, i.e., the actual aliasing in the data is less than the predicted level. Therefore, the method that predicts the least amount of aliasing is the most useful in terms of detecting the true level of aliasing. For almost all the points observed, the prediction obtained inspecting the PSD at the Nyquist limit with the assumption of monotonic decay beyond this limit is seen to be more conservative than the derivative-based prediction. The improvement using the latter method is significant similar to the case seen in the model problem in \S\ref{subsec:gldetec}. Regarding the predictions, streamwise velocity does not suffer from aliasing even very near the nozzle, while pressure is aliased at $x/D=0.1$ on the lip line. 

The LES database does not contain the time-derivative of the forcing. In that case, to predict aliasing in the forcing, we will follow the strategy described in \S\ref{subsubsec:inspect}, which assumes monotonic decay beyond the Nyquist limit. Figure \ref{fig:forcepsd} shows the PSD of the forcing terms at various stream-wise positions on the lip-line. In the downstream regions $x/D>3$, the spectra have decayed by two orders of magnitude when the Nyquist limit is reached, while the decay is less in the upstream regions and the spectra remain flat for $x/D<1$. A map showing the ratio given by \eqref{eq:aliasrat} as a measure of aliasing is presented in figure \ref{fig:forceflat} for the forcing terms corresponding to streamwise and radial momentum and energy, $f_{u_x}$, $f_{u_r}$, and $f_p$, respectively. The upper limit of the color index is set to 1 which corresponds to aliasing terms having the same magnitude as the true spectrum. For all the forcing components, predicted aliasing is seen to be larger in amplitude than the state itself for $x/D<2$ around the lip line. We also see significant aliasing within the potential core, particularly for the component $f_p$. This is due to very low forcing amplitude in this region leading to low signal-to-noise ratio. A similar phenomenon is observed for $f_p$ above the lip line. 

Forcing data around the lipline near the nozzle is critical in terms of resolvent-based response prediction. The optimal forcing mode has its entire spatial support contained in this region  \citep{towne_jfm_2018,schmidt_jfm_2018,lesshafft_prf_2019}, thus it is important to obtain accurate forcing data in this region. We emphasize that the Nyquist limit of $St=6.25$ for the database we consider is above that of many LES databases used in the literature on jets, and yet, is seen to be insufficient for calculation of forcing terms, which is a crucial step of resolvent analysis outlined in \S\ref{sec:numcase}. Increasing the sampling rate is thus impractica1 and the anti- and de-aliasing strategies proposed must be used.
\subsection{Cost of existing anti-aliasing solutions for the LES database}
\subsubsection{Increasing the sampling rate}
 The LES database we considered is sampled at $St_{s}=12.5$. A small selection of probes were available for which a higher sampling rate of $St_{s}=50$ was used. A convergence analysis was conducted by sampling this probe data at $St=50$, 25 and 12.5 and comparing the spectra of the resulting signals. The comparison is given in figure \ref{fig:probe} for $u_x$ and $p$. The Nyquist limit shown corresponds to that of the current database. It is seen that the PSD of $u_x$ remains nearly constant up to a certain frequency and shows an exponential decay beyond it. The point for exponential decay moves to lower frequencies as the probe data moves downstream of the nozzle, causing the energy contained in the signal at the Nyquist limit to sufficiently decrease, and thus, yielding an un-aliased spectrum for the entire domain. For the pressure, on the other hand, there exists a broadband peak in the PSD that reaches the Nyquist limit at $x/D=0.05$, causing significant aliasing for the entire frequency range. Similar to the case of $u_x$, the peak moves to lower frequencies with the probe moving downstream. Convergence {in the predicted spectrum}, in this case, is obtained only after $x/D=1$ for the frequency range $St=[0.1,1]$.

\begin{figure} [htb]
  \centerline{\resizebox{\textwidth}{!}{\includegraphics{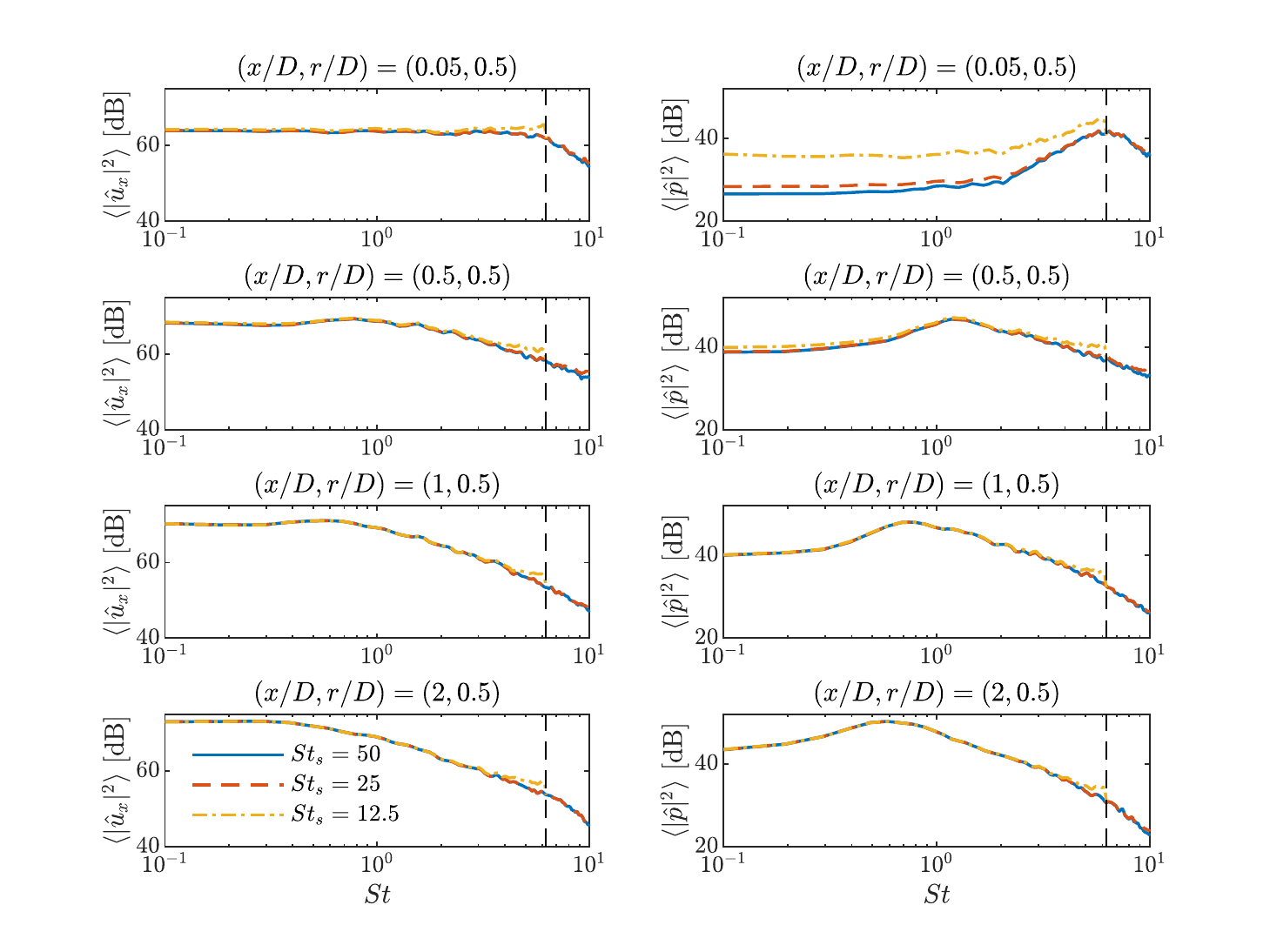}}}% Images in 100% size
  \caption{PSD of $u_x$ (left) and $p$ (right) obtained using the high time resolution data at various axial locations on the lip line. The vertical dashed lines indicate the Nyquist limit for the LES database.}
\label{fig:probe}
\end{figure}

The results show that convergence is achieved for $u_x$ with sampling rate equal to $St=25$ in all probe locations while this is the case for $p$ for all probe positions except $x/D=0.05$, i.e. nozzle exit. Once again, this region is critical for resolvent-based response prediction. As discussed in \S\ref{sec:aliasing}, the level of aliasing in the forcing terms is expected to be higher than in the state variables. Significant aliasing in $p$ at the nozzle exit thus indicates a potentially higher aliasing in the forcing terms and over a larger spatial extent. To have alias-free forcing data over the entire flow domain, sampling the data at a rate higher than $St=50$ is necessary. The current dataset requires 6.6 Tb of storage size with 2000 acoustic units of simulation time on an optimized grid of 16 million control volumes, which constitutes the minimum configuration for converged flow statistics. Increasing the sampling rate to $St=50$ would lead to a storage size of over 25 Tb. Such a large storage cost makes this approach inapplicable to more complex flow cases that would require a mesh significantly larger in size. This is yet another indication of the necessity to find optimized solutions for the aliasing issue in data-driven analyses in fluid mechanics. 
 
\subsubsection{Low-pass filtering in the time domain} \label{subsubsec:lpassfilt}
The attenuation level may not be optimally determined prior to LES analysis since the resulting spectrum would be unknown. For our test case, we will make use of the probe data to design the optimum filter with minimum order that yields at least two orders of magnitude difference between the true spectrum and the aliasing. In figure \ref{fig:probe}, a 10 dB difference between the PSD of pressure is observed at the Nyquist limit and at $St=1$. The attenuation level is then set as 30 dB, which makes the difference to be $-40$ dB, corresponding to two orders of magnitude. A `Kaiser' window FIR filter \citep{kaiser_ieee_1980} is adopted, as it minimizes the ripples. The filter is designed using Matlab's filter design tool. With the given LES time-step that corresponds to an initial sampling rate of $St=2500$, the filter order satisfying the above specifications is found to be 733. The filter order indicates the number of snapshots to be stored in the memory on the fly, which makes implementation of such a filter impracticable even for problems with a small number of degrees of freedom. Note that the attenuation level is kept at minimum thanks to readily available spectral information, which would not be the case in a general application. The order would be 1423 for the same filter if the attenuation level was set as 50 dB.

As discussed in \S\ref{subsec:lpf}, one can significantly reduce the filter order by using a cascaded filter. To avoid any interpolation of data, downsampling rates at each stage are constrained to be integers. The specifications and the resulting filter orders of a 3-stage cascaded Kaiser window FIR filter are listed in Table \ref{tab:3stageFilt}.
\begin{table}
\caption{The details of the cascade for the 3-stage Kaiser window FIR filter} \label{tab:3stageFilt}
\begin{center}
\begin{tabular}{c c c c}
Stage & $St_{samp}$ & $St_{stop\text{-}band}$ & Filter order \\\hline
1 & 2500 & 0.49*312.5 & 27 \\
2 & 312.5 & 0.49*62.5 & 18 \\
3 & 62.5 & 0.49*12.5 & 21 \\
\end{tabular}
\end{center}
\end{table}
The cumulative filter order is calculated by summing up the filter order at each stage, which gives 56. It is seen that the cascaded filter approach provides an order of magnitude reduction in the cumulative filter order compared to single-stage filter. 

The order of various cascaded FIR filters that provide 30 db attenuation at each stage while keeping the initial and final sampling rates constant are reported in Table \ref{tab:2}. 
\begin{table}
\caption{Cumulative orders of various cascaded filters providing 50 dB attenuation with an overall downsampling ratio of 200} \label{tab:2}
\begin{center}
\begin{tabular}{l | c c c c c}
\# of stages & 1 & 2 & 3 & 4 & 5 \\
Filter order & 733 & 115 & 66 & 59 & 59\\
\end{tabular}
\end{center}
\end{table}
It is seen that the filter order converges to a minimum, beyond which increasing the number of stages is not helpful. Despite an order of magnitude reduction in filter order, storing around 50 snapshots in memory for every term to be filtered may still be infeasible for large problems. This exercise highlights the need for alternative strategies.

An alternative approach applicable for problems where keeping the phase information is not necessary is to implement IIR filters, which are significantly more efficient than FIR filters. Using the same tool to design a Chebyshev type II IIR filter with the same specifications as before results in a filter order of 6. Considering that both $\mathbf{q}$ and $\partial \mathbf{q}/\partial t$ should be calculated and filtered at runtime, the total filtering cost raises to 12 snapshots. Given the size of a single snapshot for our test case $\sim$1 Gb, the cost of the IIR filter may be affordable in certain cases. However, the increase in the memory footprint of the problem makes it intractable for more complex problems where the mesh size can reach up to hundreds of millions of elements.  

\subsection{Derivative based de-aliasing}

We now test if the state data obtained from the LES can be de-aliased using the time-derivative approach. The LES database is probed at certain locations at a high sampling rate, $St=50$. The results of derivative-based de-aliasing is compared against the high-time-resolution probe data in figure \ref{fig:dealtime} assuming that the probe data provide a spectrum with negligible aliasing. The downsampled data is extracted from the LES database. It is seen that the method removes the aliasing effect to a significant extent in both $u_x$ and $p$ data, particularly at frequencies close to the Nyquist limit, $St=6.25$. This can be a desired property when the spectra decay in a monotonic fashion as in $u_x$. In that case, the frequencies close to the Nyquist limit are most affected by aliasing and can be effectively de-aliased using the derivative-based method. }

\begin{figure} [htb]
  \centerline{\resizebox{\textwidth}{!}{\includegraphics{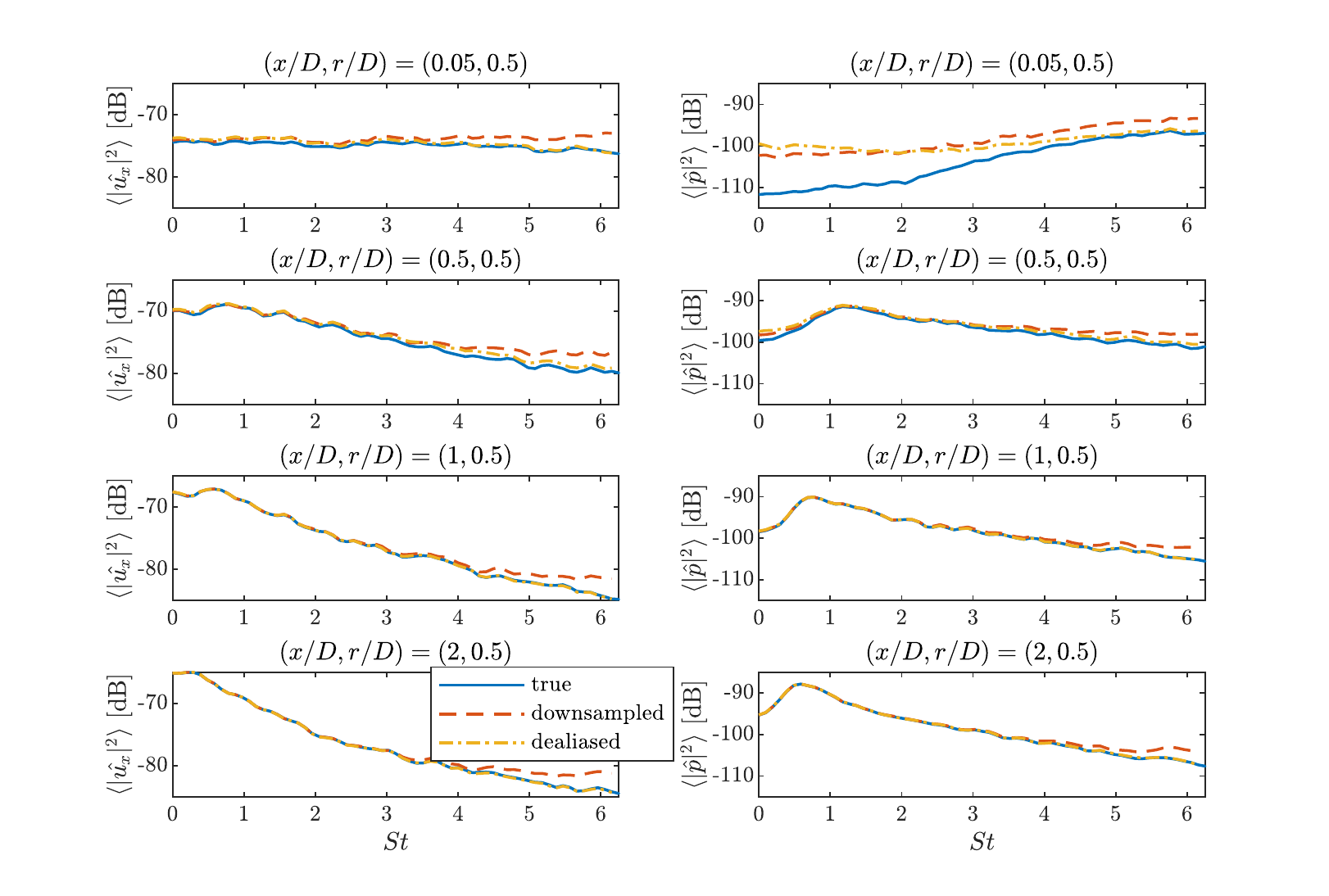}}}% Images in 100% size
  \vspace{-10px}
  \caption{Comparison of the de-aliased PSDs (solid) of $u_x$ (left) and $p$ (right) against the true spectra (dashed) and the PSDs of the downsampled spectra (dash-dotted) at various positions.}
\label{fig:dealtime}
\end{figure} 

%\begin{figure} [htb]
%  \centerline{\resizebox{\textwidth}{!}{\includegraphics{al_aliasedSpec-eps-converted-to.pdf}}}% Images in 100% size
%  \caption{Comparison of the true PSDs of $u_x$ (left) and $p$ (right) and the first three aliasing terms when sampled at $St=12.5$ at various positions. The vertical dashed line indicates the frequency where the magnitude of the true spectrum and the first aliasing terms differ by 10 dB. The vertical dash-dotted line indicates the frequency where the magnitude of the first and second aliasing terms differ by 10 dB.}
%\label{fig:aliased}
%\end{figure}

{The method is seen to remove aliasing in $u_x$ near the Nyquist limit for all the positions investigated. Same results are observed for pressure at all the positions except the near nozzle region. For the pressure near the nozzle, the de-aliasing method fails to approximate the true spectrum at low frequencies. Below a certain frequency, the difference with the true spectrum becomes even larger than that in the uncorrected data. Once again, this is due to the violation of the assumption that the leading aliasing term is dominant as discussed in \S\ref{subsec:gltd}. %Using the high-time-resolution probe data, we predict the first three aliasing terms in case of sampling rate equal to $St=12.5$. The true spectrum and the aliasing terms for $u_x$ and $p$ are shown in figure \ref{fig:aliased} at various axial positions. The vertical dash-dotted lines indicate the frequency where the difference between the PSDs of the first and second aliasing terms reach 10 dB, considered as the minimum level for dominant first aliasing assumption to be valid. Similarly, the vertical dashed lines indicate the frequency where the difference between the true spectrum and the first aliasing term reach 10 dB. For the frequency range to the right of dash-dotted line, the de-aliasing method can be used to approximate the true spectrum. For the frequency range to the left of the dashed line, aliasing can be considered negligible. For the pressure at $x/D=0.05$ on the lip line, the first aliasing term is dominant over the true spectrum, hence, there is no frequency region with negligible aliasing. Besides, below $St=4$, the first aliasing term is not dominant over the second term, which explains the poor performance of the de-aliasing method in this region. Note, however, that the pressure near the nozzle has a broad spectral content, and thus with considerable aliasing effects in it. For all other cases, the proposed approach significantly improves the PSD estimation.}

\subsection{De-aliasing by spatial filtering} \label{eq:spatfilt}
Dynamics of jets and shear flows are locally dominated by convective mechanisms. The turbulent structures are mostly convected with a velocity proportional to, and aligned with, the mean velocity. This implies that spatial filtering can be used to eliminate the high-frequency content in the data, and thus, to reduce aliasing a posteriori. To use this approach, the flow field should satisfy two criteria: it should be sufficiently refined so as not to have aliasing in the spatial distribution, and the statistical description of the flow field should remain nearly constant within the spatial support of the filter. The structured cylindrical grid is designed to have a resolution in the axial and radial directions similar to that of the LES grid. Given the slowly varying nature of the jet, it is possible to define a window over which the spectral content of the flow remains nearly constant for most of flow field. To check the validity of this assumption and to determine the characteristics of the spatial filter to be applied, the wavenumber-frequency spectra of the flow is plotted in figure \ref{fig:wavefreq}. As the level of aliasing varies along the jet axis, we investigate the wavenumber-frequency spectra at multiple axial stations to better evaluate the evolution of the aliasing. 

\begin{figure} [htb]
  \centerline{\resizebox{0.85\textwidth}{!}{\includegraphics{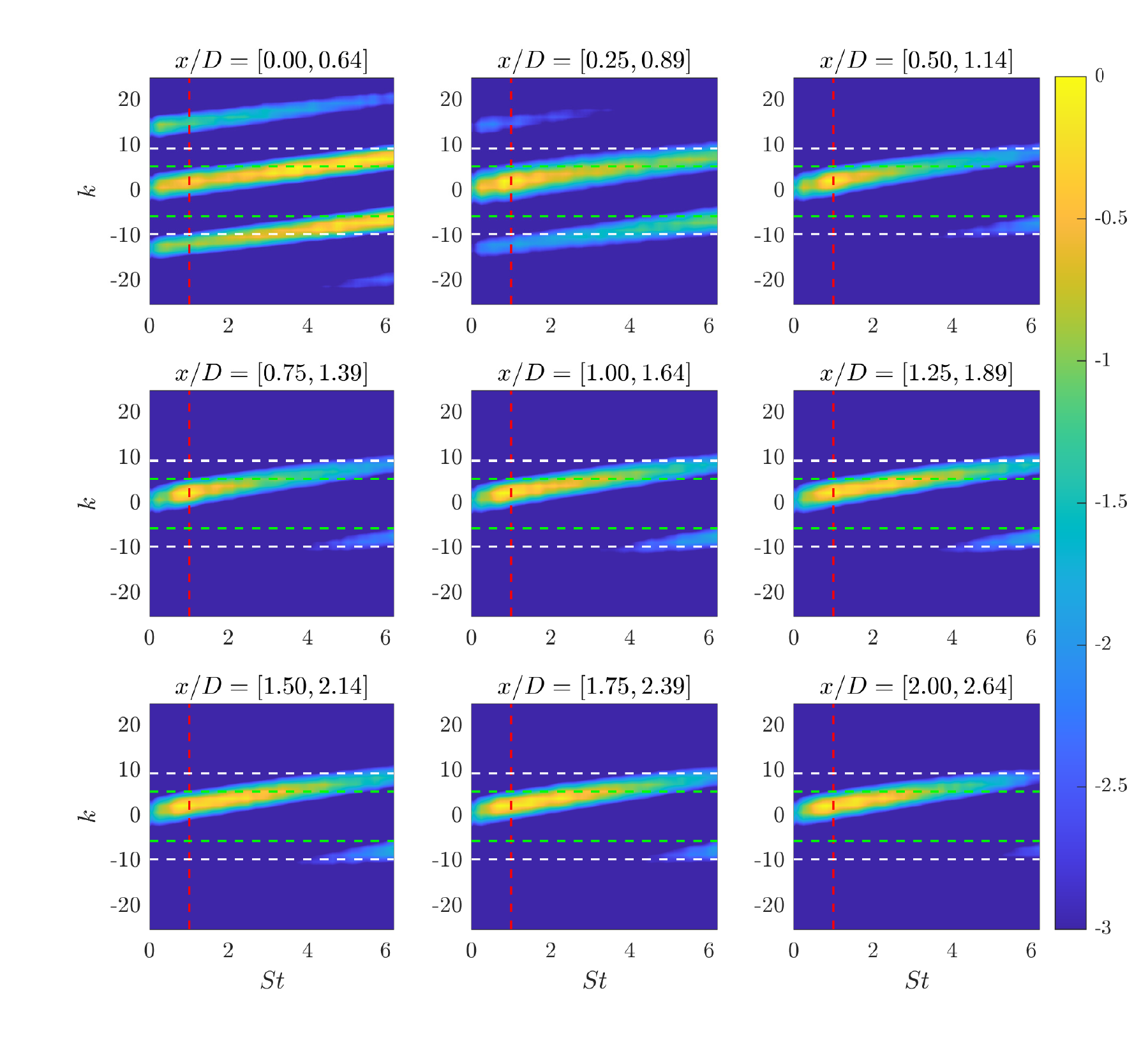}}}% Images in 100% size
  \vspace{-10px}
  \caption{Wavenumber-frequency spectrum of $f_{u_x}$ at various axial segments of length equals $0.64D$ below the lip line ($r/D=0.4$). The vertical-dashed-red lines indicate the maximum frequency of interest. The horizontal-dashed-green and -white lines indicate the cut-off and the stop-band wavenumbers of the spatial filter, respectively.}
\label{fig:wavefreq}
\end{figure}

The wavenumber domain is calculated in the axial direction. The data at a given radial position is interpolated onto an equispaced grid in the axial direction with size equal to the minimum grid size in the structured grid. The number of points used to compute a discrete-space Fourier transform in the $x$-direction is set as 128 yielding a window length of 0.64 $D$. A hanning window is used to reduce spectral leakage. Note that the window length is to be kept to a minimum in view of the assumption of constant statistics over the filter support. The wavenumber-frequency spectra calculated for $f_{u_x}$ at various positions are shown in figure \ref{fig:wavefreq}. The spectra contain separate bands of energy in the wavenumber-frequency domain, which are indeed pieces of a single band chopped due to the Nyquist limit of the frequency axis. Similar to the spectra shown in figure \ref{fig:spatfilt} for the model problem, these bands except the one in the center represent the aliasing terms. 

The vertical-dashed-red line indicates the maximum frequency of interest for our test case. A spatial filter applied on the flow field acts in the wavenumber direction. The horizontal, dashed-green and -white lines indicate, respectively, the end of the pass-band and the beginning of the stop-band of the spatial filter. The limits of the spatial filter are selected such that the true spectra remain within the pass-band while all the aliasing terms remain in the stop-band for the frequency range of interest. Keeping the pass-band and the stop-band lines too close to each other increases the filter length, while a shorter filter is preferred as mentioned above. An FIR type spatial filter yielding 30 dB attenuation between the cut-off and stop-band wavenumbers, in the given case, equal to 5.5 and 9.5, respectively, is applied on the data. The resulting filter order is 79 corresponding to a filter length of 0.4 $D$. The spectra presented in figure \ref{fig:wavefreq} indicate that, beyond a certain frequency, $St>\sim2$ in the present case, any spatial filter that attenuates all the aliasing terms is bound to attenuate part of the true spectrum as well. Therefore one may have an upper limit of frequency depending on the wave-number spectra, for the spatial filtering to be a valid tool to remove aliasing. A similar issue was observed \S\ref{subsec:glspat} when de-aliasing database in the model problem. Figure \ref{fig:wavefreq} also shows that shifting the interrogation window near the nozzle by 0.25 $D$, which corresponds to nearly half of filter length, yields a significant change in the resulting wavenumber-frequency spectrum. Therefore, the assumption of flow with constant spectral content within the filter length is not fully satisfied in this region, which may again limit the validity of the technique.

The effect of the spatial filter on the frequency spectrum of the forcing terms corresponding to streamwise momentum and energy are compared to the unfiltered spectra at various locations in figure \ref{fig:spatFiltEffect}. It is seen that, for $x/D>1$, the filter can deliver $\sim$20 dB decay before reaching the Nyquist limit, while for near the nozzle, the spectrum remains flat, indicating aliasing. This shows that with spatial filtering, it is possible to remove aliasing from the existing database for a vast majority of the flow field. 

\begin{figure} [htb]
  \centerline{\resizebox{0.75\textwidth}{!}{\includegraphics{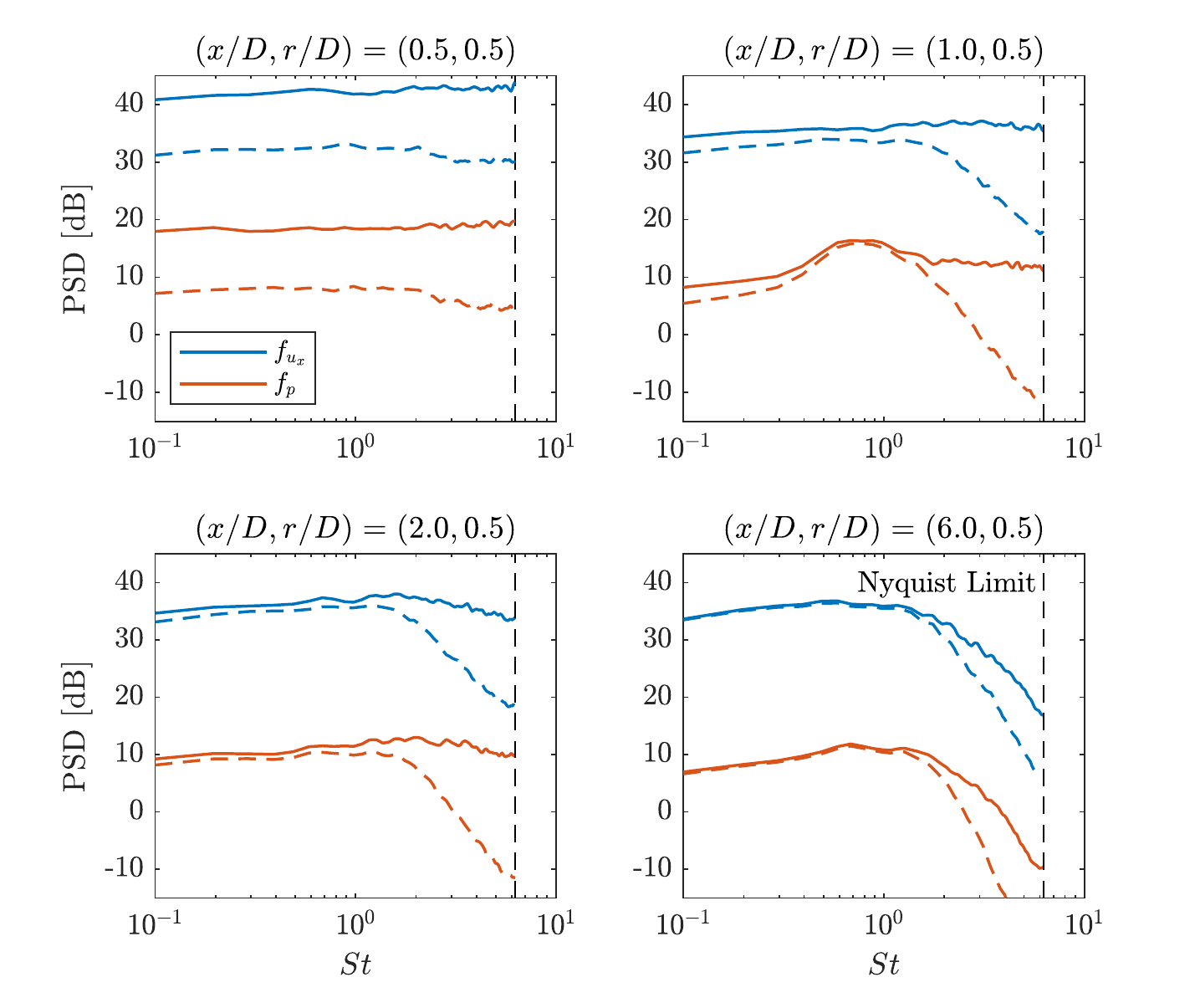}}}% Images in 100% size
  \caption{Comparison of the spatially filtered spectra (dashed) of the forcing terms $f_{u_x}$ and $f_p$ (blue and orange, respectively) against the unfiltered spectra (solid) at various positions. Dashed and solid lines indicate the original and the spatially filtered data, respectively. The vertical dashed line indicates the Nyquist limit.}
\label{fig:spatFiltEffect}
\end{figure}

\section{Conclusions}
We discussed treatment of aliasing in large databases where classical solutions can be infeasible due to large computational or storage costs. Typical solutions to the aliasing problem, such as increasing the sampling rate or low-pass filtering the data a priori, are generic in the sense that they can be applied to any time-dependent system without knowledge of its dynamical properties, but, as shown here, are often impractical for use in large flow databases. In this paper, we proposed strategies to detect and mitigate aliasing in dynamical systems defined by a set of governing equations, taking advantage of certain characteristics associated with the systems, reducing the added computational cost and database size. We investigated aliasing in both the state and the nonlinear terms that appear in the Navier-Stokes (N-S) equations. When using a resolvent formulation of the N-S equations, the nonlinear terms appear as a forcing that drives the mean-flow-based resolvent operator to generate the state. {Having an accurate spectral representation of the forcing term is known to be critical in certain flow cases for resolvent-based modelling \citep{karban_jfm_2020,nogueira_jfm_2021,morra_jfm_2021}, estimation \citep{towne_jfm_2020,martini_jfm_2020}, and control \citep{martini_jfm_2022}.}

{A simple way to detect aliasing in a signal is to analyse the spectral decay near the Nyquist limit. Assuming a monotonic decay in the spectrum beyond the Nyquist limit, the ratio of the PSD value obtained at a given frequency to the value obtained at the Nyquist limit provides a prediction of aliasing at that frequency. We showed that using the time-derivative information, one can predict the aliasing with improved accuracy. For simulation data, the time-derivative of the state can be obtained from the governing differential equations even after downsampling.}

{The common practice for anti-aliasing a signal before downsampling is to apply a low-pass filter to satisfy the Nyquist criterion. We discussed different low-pass filter types to be used for anti-aliasing. A multi-stage low-pass filtering approach was proposed as a means to reduce the filter order, and thus, the cost of the filter. For a high sampling rate, it was shown that the multi-stage filter reduces the filtering cost by an exponential factor of the number of stages.}

{ We introduced several strategies for de-aliasing and anti-aliasing the state and/or the forcing data in a flow database. For de-aliasing an already downsampled database, we proposed two methods based on time-derivative information and spatial low-pass filtering, respectively. The time-derivative-based method involves assuming a single dominant aliasing term and using the time-derivative data to turn the ill-posed problem of aliasing to a well-posed linear problem yielding a prediction of the true spectrum. Assuming monotonic decay beyond the Nyquist limit, the method is expected to yield better predictions at frequencies close to the Nyquist limit. The de-aliasing method based on spatial low-pass filtering is applicable for convective systems for which a strong correlation is found between the space and time axes. It was shown that for such systems, it is possible to attenuate the high-frequency content of the data, and thus mitigate aliasing, by applying a spatial low-pass filter in the direction of convection velocity. For filtering in space to be valid, the data should be statistically homogeneous in the direction of filtering over the width of the filter stencil. In case of spatially developing flows such as boundary layers or jets, the size of the filter must be small enough to satisfy the homogeneous flow assumption. This limits the maximum attenuation that can be achieved in the wavenumber space. The method also requires that the domain is discretised with sufficient refinement to avoid spatial aliasing. Given the two de-aliasing methods, we discussed that the former is more suitable for the state data, as, once again, it is possible to compute the time derivative of the state using the governing equations, while the latter can be applied both for the state and the forcing provided that they are dominated by convective mechanisms.}

{Finally, we introduced an anti-aliasing method specific to the forcing data. The method involves storing the state and its time integral and using these to compute the forcing in the frequency domain. We discussed that the method could be effective for anti-aliasing at low frequencies. We also showed that the attenuation that can be achieved using this approach could be enhanced by applying a multi-stage integration, which yields a similar result as in the case of multi-stage low-pass filtering discussed above, but at reduced computational cost.}

{We showcased the detecting, de-aliasing and anti-aliasing methods using a model problem based on the Ginzburg-Landau equation and an LES database that is specifically designed for calculating the forcing terms in a subsonic jet. Tests on the model problem showed that derivative-based de-aliasing can eliminate aliasing at frequencies near the Nyquist limit while spatial filtering is more effective at low frequencies. These two methods, when used together, can therefore help substantially remove aliasing effects from a database in the entire frequency range. }

{Anti-aliasing using integration has been shown to significantly eliminate the aliasing in the power spectral density of the forcing reaching up to 15 dB in the model problem. The method yielded better results at lower frequencies as the attenuation is inversely proportional to the frequency. We also showed that using a multi-stage integration approach increased the attenuation. A linear increase in the number of integration stages applied is rewarded with a geometric increase in the attenuation level.}

{The size of the LES database used for benchmarking precluded applying standard techniques to remove aliasing. Using a standard FIR low-pass filtering to reduce aliasing yields a filter order $\sim$600, which corresponded to $\sim$600 Gb of storage on the fly to apply the filter. Using the multi-stage approach, the filter order was reduced by an order of magnitude. The database contained the simulation data for the state variables and their time derivatives. Comparison of state and time derivative data was used to detect the zones affected by aliasing in the flow domain. State data in the near-nozzle region was seen to contain significant aliasing. The ratio method was used to detect the aliasing in the forcing term, as no time-derivative data was available for the forcing. Aliasing in the forcing was found to be higher compared to the state, which was expected due to the fact that the forcing contains more energy at higher frequencies as a result of triadic interactions. }

{The forcing in the jet is convected with the turbulence. This allowed us to use spatial filtering in the streamwise direction to reduce the aliasing in the forcing data. The filtering provided significant improvement in predicting the true spectrum at low frequencies in the entire flow field except the shear zone near the nozzle. For the state data, the applicability of the derivative-based de-aliasing was tested. It was seen that the method could be useful for the frequencies near the Nyquist limit.} %The method see A low-pass filtering method based on integral quantities was suggested as an optimum approach to suppress the high-frequency content in the data with minimum implementation and storage costs. In the presented filter, a linear increase in the filter order was awarded by an exponential increase in the attenuation level, which was not the case when using FIR filters. 

All the methods discussed in this paper have certain constraints, mostly due to limited computational resources. They are seen to start failing near the nozzle, which is known to be critical to describe jet dynamics \citep{schmidt_jfm_2018,lesshafft_prf_2019}. One may have to choose a combination of these methods to eliminate/avoid aliasing, particularly around the shear layer near the nozzle. Although the approaches discussed here were demonstrated for a jet, extension to other problems involving high-fidelity simulation data is straightforward. 

\section*{Acknowledgements}
This study has been funded by the Clean Sky 2 Joint Undertaking under the European Union’s Horizon 2020 research and innovation programme under grant agreement No 785303. U.K. has received funding from TUBITAK 2236 Co-funded Brain Circulation Scheme 2 (Project No: 121C061). A.T. was supported in part by ONR grant N00014-22-1-2561. The LES study was supported by NAVAIR SBIR project, under the supervision of Dr J. T. Spyropoulos. The main LES calculations were carried out on CRAY XE6 machines at DoD HPC facilities in ERDC DSRC.
\section*{Data Availability Statement}
The datasets generated and/or analysed during the current study are available from the corresponding author on reasonable request.

\appendix
\section{Spatial aliasing in the LES database} \label{ap:convVel}

The present data set calculated on an unstructured grid is mapped onto a cylindrical grid to compute the FT in the azimuthal direction in a robust manner. The dimensions of the structured grid should normally be determined to provide a resolution similar to that of the LES grid in order to avoid spatial aliasing. The original grid was refined near the nozzle in the axial direction, and around the shear layer in the radial and azimuthal directions \citep{bres_aiaa_2015}. The azimuthal refinement at the shear layer brings excessive interpolation around the jet axis and causes an increase in storage cost without any benefit. A solution to this problem is to determine the minimum number of points in azimuth to map the flow data onto the structured grid without aliasing. In previous studies in which the present dataset had been used to investigate the state variables, 128 points in azimuth had been used. A convergence analysis for aliasing in the azimuthal direction for both the state variables and the forcing terms around the shear layer is conducted using a single snapshot and is depicted in figure \ref{fig:azim128}. The analysis reveals that aliasing in the state variables is negligible for the first azimuthal mode, while even the $m=0$ mode is aliased for the forcing terms. The same test is repeated on a cylindrical grid with 512 points in azimuth, which roughly corresponds to the number of points around the shear layer near the nozzle in the unstructured LES grid (see figure \ref{fig:azim512}). It is seen that for the forcing terms near the nozzle, convergence is obtained for $f_{u_x}$ with 256 points while it is not the case for $f_p$. This shows that the LES grid resolution should be kept around the shear layer near the nozzle to avoid spatial aliasing in the azimuthal direction, but at the cost of quadrupling the size of the database. 

\begin{figure} [htb]
  \centerline{\resizebox{0.85\textwidth}{!}{\includegraphics{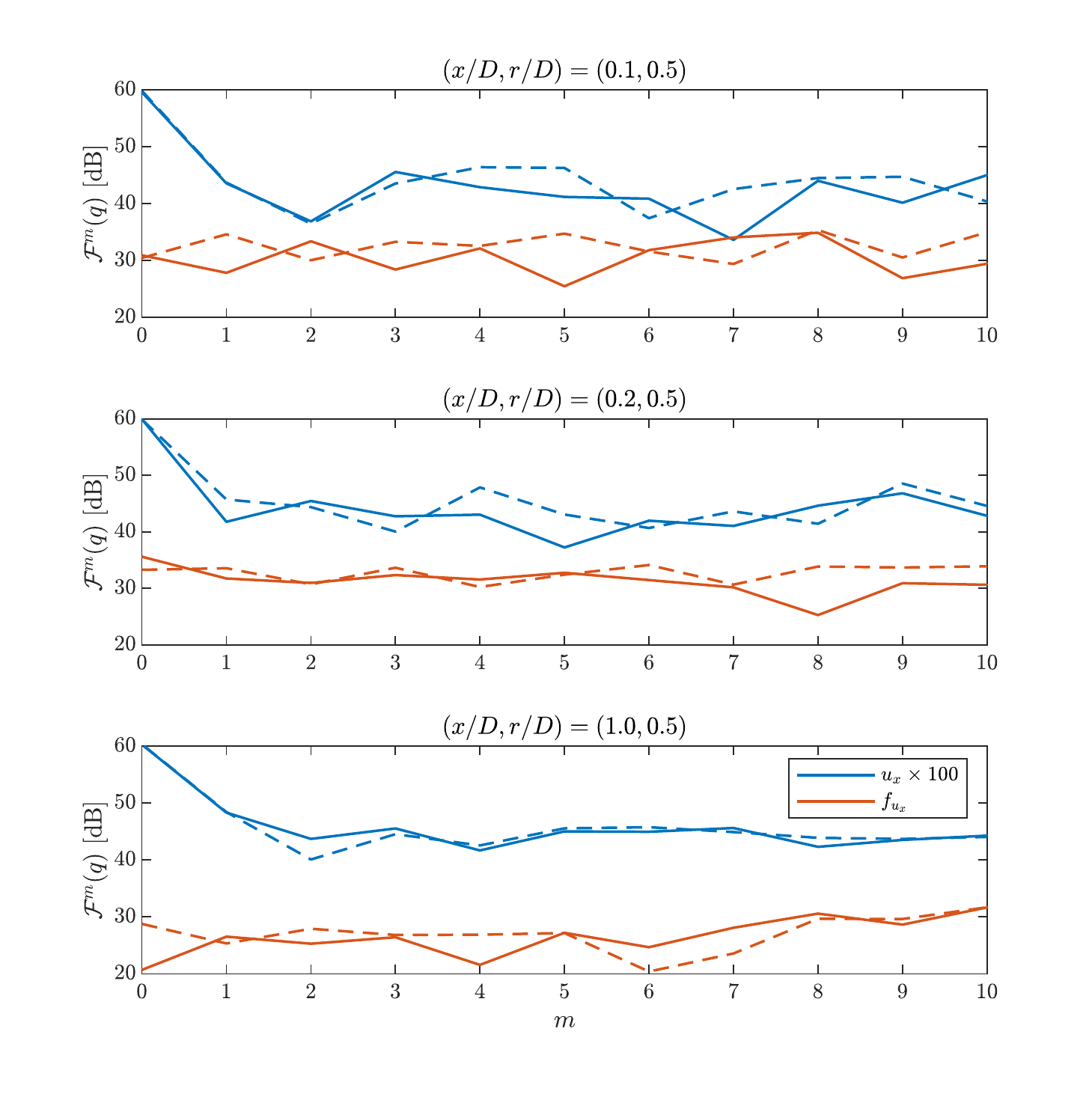}}}% Images in 100% size
  \caption{Azimuthal FT of $u_x$ (blue) and $f_{u_x}$ (orange) terms calculated at various axial positions on the lip-line. Solid and dashed lines correspond to 128 and 64 grid points, respectively, in azimuthal direction. $u_x$ is scaled by a random factor to increase readability.}
\label{fig:azim128}
\end{figure}

\begin{figure} [htb]
  \centerline{\resizebox{0.85\textwidth}{!}{\includegraphics{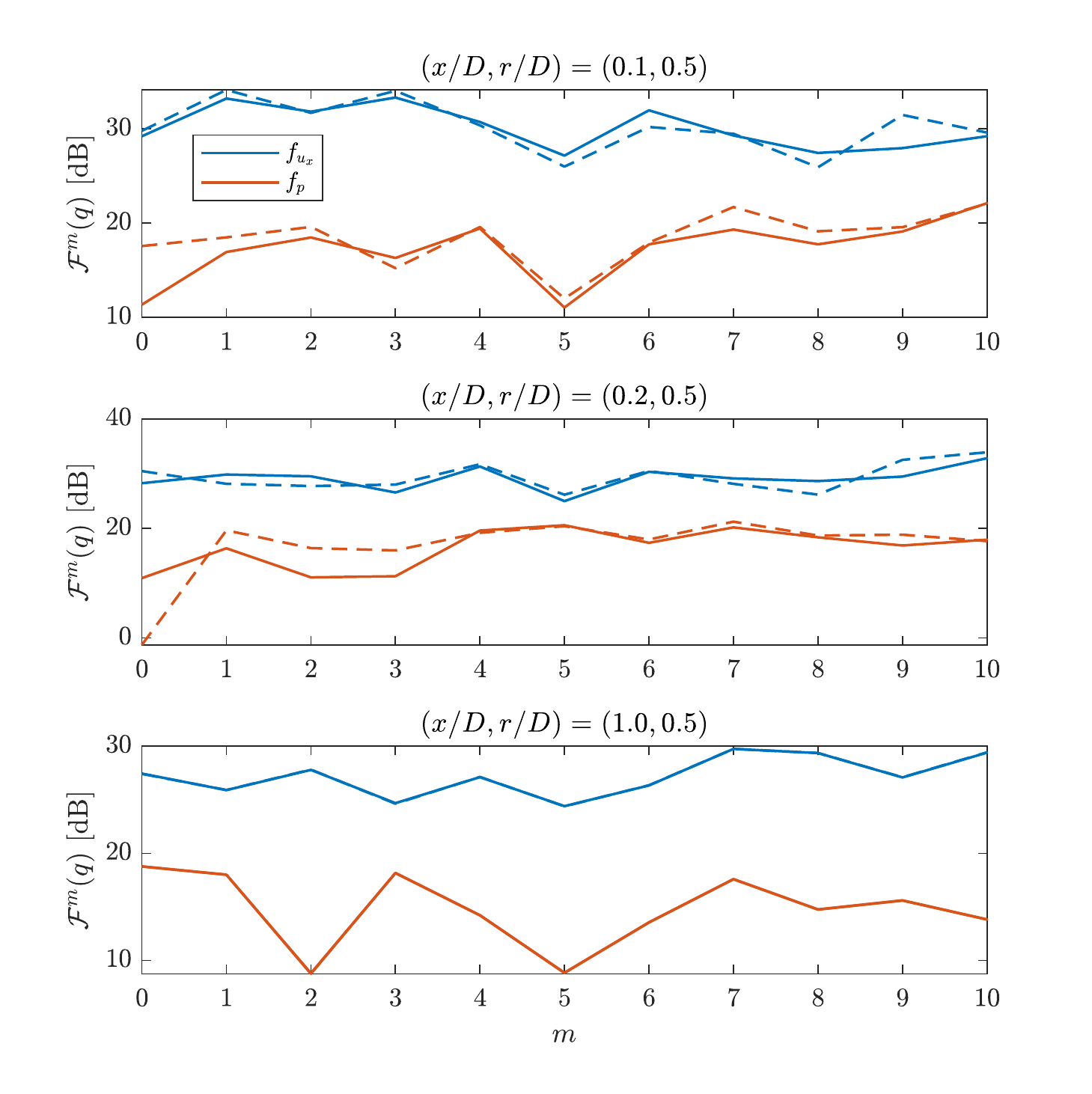}}}% Images in 100% size
  \caption{Azimuthal FT of $f_{u_x}$ (blue) and $f_{p}$ (orange) terms calculated at various axial positions on the lip-line. Solid and dashed lines correspond to 512 and 256 grid points, respectively, in azimuthal direction.}
\label{fig:azim512}
\end{figure}

An alternative solution to the mapping problem is to low-pass filter the flow field in the azimuthal direction, and to downsample afterwards obeying the Nyquist criterion. Azimuthal filtering can be applied either through a weighted moving-average filter, or by taking the azimuthal FT of the data, setting the mode numbers to be filtered to zero, and taking the inverse azimuthal FT. Since the data is already periodic, taking the FT does not cause any spectral leakage. Note that once filtered in the azimuthal direction, a direct conversion of the velocity field from Cartesian to cylindrical, or vice versa, is not valid any more since the conversion is not linear in the azimuthal direction. However, one can switch between the two velocity fields after taking the azimuthal FT. 

Conversion of velocity from Cartesian to cylindrical coordinate system is performed as 
	\begin{align}
	u_r &= u_y \cos(\theta) + u_z \sin(\theta), \label{eq:ur}\\
	u_{\theta} &= -u_y\sin(\theta) + u_z\cos(\theta), \label{eq:uth}
	\end{align}
where $\theta$ is measured from the $y$-axis. Taking the Fourier transform (FT) of \eqref{eq:ur} and \eqref{eq:uth} in $\theta$ yields the following convolution expressions:
	\begin{align}
	\hat{u}_r^{(i)} &= \hat{u}_y^{(i)}*\mathcal{F}\left(\cos(\theta)\right) +
	\hat{u}_z^{(i)}*\mathcal{F}\left(\sin(\theta)\right), \label{eq:uhatr} \\
	\hat{u}_{\theta}^{(i)} &= -\hat{u}_y^{(i)}*\mathcal{F}\left(\sin(\theta)\right) +
	\hat{u}_z^{(i)}*\mathcal{F}\left(\cos(\theta)\right), \label{eq:uhatth}
	\end{align}
where the superscript $(i)$ denotes the azimuthal mode number. Using Matlab's convention for FT, the FTs of $\cos(\theta)$ and $\sin(\theta)$ are given as
	\begin{align}
	\mathcal{F}\left(\cos(\theta)\right) &= \frac{1}{2}\left(\delta(i-1)+\delta(i+1)\right), \label{eq:cosft} \\
	\mathcal{F}\left(\sin(\theta)\right) &= \frac{i}{2}\left(\delta(i-1)-\delta(i+1)\right). \label{eq:sinft}
	\end{align}
Then the convolution expressions given in \eqref{eq:uhatr} and \eqref{eq:uhatth} can be re-written as
	\begin{align}
	\hat{u}_r^{(i)} &= \frac{1}{2}\left(\hat{u}_y^{(i-1)} + \hat{u}_y^{(i+1)}\right) +
	\frac{i}{2}\left(\hat{u}_z^{(i-1)} - \hat{u}_z^{(i+1)}\right), \label{eq:uhatr2} \\
	\hat{u}_{\theta}^{(i)} &= -\frac{i}{2}\left(\hat{u}_y^{(i-1)} - \hat{u}_y^{(i+1)}\right) +
	\frac{1}{2}\left(\hat{u}_z^{(i-1)} + \hat{u}_z^{(i+1)}\right), \label{eq:uhatth2}
	\end{align}
Similarly, conversion from cylindrical to Cartesian coordinates can be achieved using
	\begin{align}
	\hat{u}_y^{(i)} &= \frac{1}{2}\left(\hat{u}_r^{(i-1)} + \hat{u}_r^{(i+1)}\right) -
	\frac{i}{2}\left(\hat{u}_{\theta}^{(i-1)} - \hat{u}_{\theta}^{(i+1)}\right), \label{eq:uhaty} \\
	\hat{u}_z^{(i)} &= \frac{i}{2}\left(\hat{u}_r^{(i-1)} - \hat{u}_r^{(i+1)}\right) +
	\frac{1}{2}\left(\hat{u}_{\theta}^{(i-1)} + \hat{u}_{\theta}^{(i+1)}\right). \label{eq:uhatz}
	\end{align}
Once the azimuthal modes of the velocity have been calculated in the transformed coordinates, one can perform an inverse FT in $\theta$ to reconstruct the filtered velocity field.

%
% BibTeX users please use
\bibliographystyle{apalike}
\bibliography{biblio}
%\begin{thebibliography}{}
%\bibitem[Author(year)]{RefJ}
%\textbf{Author}, Journal \textbf{Volume}, (year) page numbers.
%\bibitem[Author(year)]{RefB}
%\textbf{Author}, \textit{Book title} (Publisher, place year) page numbers
% etc
%\end{thebibliography}

\end{document}